\documentclass[preprint2]{aastex}

\usepackage{amsmath,amssymb,amsfonts} 
\usepackage{color} 
\usepackage[bookmarks=true]{hyperref}
\hypersetup{
colorlinks=true,
pdfborder={0 0 0},
linkcolor=red,
citecolor=blue
 }
\usepackage{natbib}
\newcommand \beq {\begin{equation}}
\newcommand \enq {\end{equation}}

\newcommand \rhod  {\rho_{\rm d}}
\newcommand \rhog  {\rho_{\rm g}}
\newcommand \rhogi  {\rho_{\rm g0}}
\newcommand \rhodi  {\rho_{\rm d0}}

\newcommand \rhoone {\rho^\ast_{\rm I }}
\newcommand \rhotwo {\rho^\ast_{\rm II}}
\newcommand \sech  {\mathrm{sech}\,}

\newcommand \Hdagger {H^\dagger_{\rm g}}
\newcommand \Hd {H_{\rm d}}
\newcommand \Hg {H_{\rm g}}
\newcommand \Qg {Q_{\rm g}}
\newcommand \Qd {Q_{\rm d}}
\newcommand \Qdtwod {Q_{\rm d, 2D}}
\newcommand \cg {c_{\rm g}}
\newcommand \cd {c_{\rm d}}
\newcommand \zd {z_{\rm d}}
\newcommand \ts {t_{\rm stop}}
\newcommand \metal {\Sigma_{\rm d}/\Sigma_{\rm g}}
\newcommand \Sigmad {\Sigma_{\rm d}}
\newcommand \Sigmag {\Sigma_{\rm g}}

\slugcomment{Manuscript in preparation}

\shorttitle{Planetesimal Formation}
\shortauthors{Shi \& Chiang}

\begin{document}


\title{From Dust to Planetesimals: \\Criteria for Gravitational Instability of Small Particles in Gas}


\author{Ji-Ming Shi\altaffilmark{1,2} and Eugene Chiang\altaffilmark{1,2,3}}

\altaffiltext{1}{Department of Astronomy, UC Berkeley, Hearst Field Annex B-20,
    Berkeley, CA 94720-3411}
\altaffiltext{2}{Center for Integrative Planetary Science, UC Berkeley, Hearst Field Annex B-20,
    Berkeley, CA 94720-3411}
\altaffiltext{3}{Department of Earth and Planetary Science, UC Berkeley, 307 McCone Hall,
    Berkeley, CA 94720-4767}
    
\email{jmshi@berkeley.edu}

\begin{abstract}
  Dust particles sediment toward the midplanes of protoplanetary
  disks, forming dust-rich sublayers encased in gas. What densities
  must the particle sublayer attain before it can fragment by
  self-gravity?  We describe various candidate threshold densities.
  One of these is the Roche density, which is that required for a
  strengthless satellite to resist tidal disruption by its primary.
  Another is the Toomre density, which is that required for
  de-stabilizing self-gravity to defeat the stabilizing influences of
  pressure and rotation.  We show that for sublayers containing
  aerodynamically well-coupled dust, the Toomre density exceeds the
  Roche density by many (up to about 4) orders of magnitude.  We
  present 3D shearing box simulations of self-gravitating, stratified,
  dust-gas mixtures to test which of the candidate thresholds is
  relevant for collapse. All our simulations indicate that the
  larger Toomre density is required for collapse. This result is
  sensible because sublayers are readily stabilized by pressure.
  Sound-crossing times for thin layers are easily shorter than
  free-fall times, and the effective sound speed in dust-gas
  suspensions decreases only weakly with the dust-to-gas ratio (as the
  inverse square root). Our findings assume that particles are small
  enough that their stopping times in gas are shorter than all other
  timescales. Relaxing this assumption may lower the threshold for
  gravitational collapse back down to the Roche criterion. In
  particular, if the particle stopping time becomes longer than the
  sound-crossing time, sublayers may lose pressure support and
  become gravitationally unstable.

\end{abstract}


\keywords{hydrodynamics --- instabilities --- planets and satellites: formation --- protoplanetary disks --- methods: numerical}

\section{INTRODUCTION \label{sec:intr}}

Gravitational instability is an attractive mechanism to form
planetesimals, but how it is triggered in protoplanetary disks remains
unclear. In one proposed sequence of events, most of the disk's solids
first coagulate into particles 0.1--1 m in size at orbital distances of a few AU. These
``boulder''-sized bodies then further concentrate by the aerodynamic
streaming instability (\citealt{youdingoodman05};
\citealt{johansenetal07}; \citealt{baistone10}; and references
therein).  Local densities are so strongly enhanced by the streaming instability
that they can exceed the Roche density (see \S\ref{sec:sekiya}
for a definition),
whereupon
collections of boulders may undergo gravitational collapse into more
massive, bound structures.

\begin{figure}[h!]
\epsscale{0.9}
\plotone{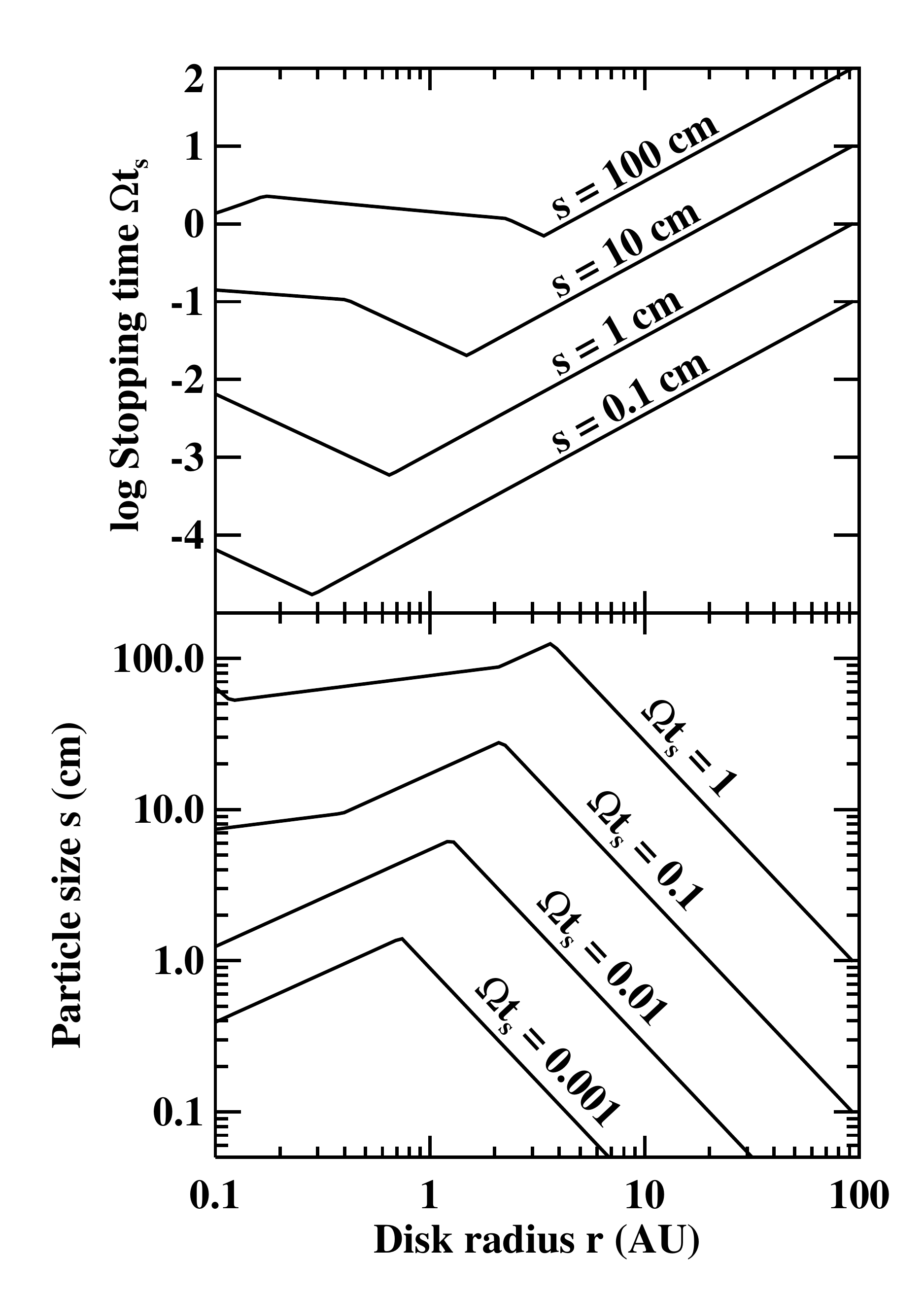} \caption{\small{Stopping times of particles in the MMSN, normalized to the local dynamical time $\Omega^{-1}$. Disk parameters are taken from \citet{chiangyoudin10}. Particles are assumed spherical with bulk density 1 g/cm$^3$. The kinks in the curves are due to transitions between different drag force laws as taken from Weidenschilling (\citeyear{weidenschilling77}; note that the transition between the Stokes and Epstein drag laws occurs when the gas mean free path equals 2/9 of the particle radius, not 4/9 as misprinted in that article). Marginally coupled particles ($\Omega \ts \sim 1$) correspond to meter-sized boulders at $r \sim 1$ AU; decimeter-sized rocks at $r \sim 10$ AU; and cm-sized pebbles at $r \sim 100$ AU. The top panel plots $\Omega \ts$ at various fixed particle radii $s$; the bottom panel plots the same data but at fixed $\Omega \ts$. In this paper we are interested in the small particle $\Omega \ts \ll 1$ limit.} }\label{fig:tstop}
\end{figure}

A weakness of this scenario is that it presumes that particle-particle
sticking (i.e., chemical adhesion) can convert most of the disk's
solids into boulders, or more accurately, particles whose momentum
stopping times in gas
\begin{equation}
t_{\rm stop} \equiv \frac{mv_{\rm rel}}{F_{\rm drag}}
\end{equation}
are within a factor of 10 of the local dynamical time $\Omega^{-1}$,
where $\Omega$ is the Kepler orbital frequency, $m$ is the particle
mass, $v_{\rm rel}$ is the relative gas-particle velocity, and $F_{\rm
  drag}$ is the drag force whose form varies with disk environment
(see, e.g., \citealt{weidenschilling77}).  Figure \ref{fig:tstop}
relates $t_{\rm stop}$ to particle radius $s$ as a function of disk
radius $r$ in the minimum-mass solar nebula (MMSN).
For $r \approx 1$--10 AU, the condition $\Omega
t_{\rm stop} = 0.1$--1 corresponds to $s \approx 0.1$--1 m. 

Unfortunately, particle-particle sticking might not produce boulders
in sufficient numbers for the streaming instability to be significant.  A comprehensive
study by Zsom et al.~(\citeyear{zsom11}; see also
\citealt{birnstiel10}) found that for realistic, experiment-based
sticking models that include both bouncing and fragmentation,
particles no larger than $\sim$1 cm can form by sticking --- even when
the disk is assumed to have zero turbulence. According to Table 1 of
\citet{zsom11}, coagulation models over most of parameter space
produce $\tau_s \sim 10^{-4}$--$10^{-2}$. This range is too small for
the streaming instability to concentrate particles strongly---see \citet{baistone10}, who showed
that when half or more of the disk's solid mass has $\Omega t_{\rm
  stop} < 0.1$, densities enhanced by the streaming instability still fall short of the Roche density
by more than a factor of 10. Even if particle-particle sticking could
grow bodies with $\Omega t_{\rm stop} \sim 0.1$--1 (e.g.,
\citealt{okuzumi09}, who neglected fragmentation), the disk's solids
may not be transformed into such bodies all at once. Rather, boulders
may initially comprise a minority on the extreme tail of the size
distribution. Unless they can multiply from a minority to a majority
within the time it takes for them to drift radially inward by gas drag
($\sim$100--1000 yr starting at 1 AU; \citealt{weidenschilling77}),
they threaten to be lost from the nebula by drag.

We are therefore motivated to ask whether gravitational instability is
practicable for particles having realistically small sizes and
concomitantly short stopping times, say $\Omega t_{\rm stop} \lesssim
10^{-2}$. Smaller particles suffer the disadvantage that they are
harder to concentrate; since they are well-entrained in gas,
turbulence in the gas can loft particles above the midplane and
prevent them from collecting into regions of higher density. 
The streaming instability provides one source of turbulence. Another driver
of turbulence is the Kelvin-Helmholtz
instability, caused by vertical velocity gradients which steepen as
dust settles into a thin, dense ``sublayer'' at the disk midplane
\citep{weidenschilling80}. Several recent studies (\citealt{chiang08},
\citealt{leeetal10,leeetal10b}; see also
\citealt{weidenschilling06,weidenschilling10}) have measured the
maximum sublayer densities permitted by the Kelvin-Helmholtz
instability. Neglecting self-gravity, they found that dust-to-gas
ratios between $\sim$2--30 are possible in disks that are locally
enriched in metallicity by factors of 1--4 above solar. Such local
enrichment can be generated by radial drifts of particles relative to gas
(see \citealt{chiangyoudin10} for a review). For
observational evidence of radial segregation of dust from gas,
see \citet{andrewsetal12}.

Are such enhancements in the local dust-to-gas ratio sufficient to
spawn planetesimals?  How high must dust + gas densities be before the
effects of self-gravity manifest?  Our paper addresses these questions
in the limit $\Omega t_{\rm stop} \ll 1$, i.e., in the limit that
particles are small enough to be well coupled to gas. In the next two
subsections, we derive critical densities for gravitational
instability in the cases of a pure gas disk (\S\ref{sec:intro_gas}),
and a disk composed of both gas and perfectly entrained ($\Omega
t_{\rm stop} \rightarrow 0$) dust (\S\ref{sec:intro_mixed}).  The two
cases give remarkably different answers for dust-rich sublayers. In \S\ref{sec:sekiya}
we add two more densities from the literature to the list of proposed criteria for
gravitational collapse. Table \ref{tab:tab1}
summarizes the various candidate threshold densities.

\begin{deluxetable}{ccll}
\tabletypesize{\footnotesize}
\tablecolumns{4} \tablewidth{0pc}
\tablecaption{Candidate Critical Densities for Gravitational Collapse}
\tablehead{\colhead{\begin{tabular}{c} Critical \\  Density \end{tabular}} & 
           \colhead{\begin{tabular}{c} Value    \\          \end{tabular}} &
           \colhead{\begin{tabular}{c} Comment  \\          \end{tabular}} & 
	   \colhead{\begin{tabular}{c} Reference\\          \end{tabular}}}
\startdata
$\rhoone$  & $\frac{1}{2\pi}\frac{1}{\Qg^\ast} \frac{\Hg^\dagger}{\Hg} \frac{M_\ast}{r^3} \sim 0.16 \frac{M_\ast}{r^3}$ \tablenotemark{(a)}
                       & \begin{tabular}{p{5cm}}
		         Equivalent to $\Qg<\Qg^\ast$ for pure gas disks 
                         \end{tabular}
                       & \begin{tabular}{p{4cm}}
		         This paper, equation (\ref{eqn:rhoI})  
			 \end{tabular} \\
\tableline
$\rho^\ast_{\rm Sekiya}$  & $0.60 \frac{M_\ast}{r^3}$ 
                       &\begin{tabular}{p{5cm}} 
		        Required for the onset of an incompressible, axisymmetric overstable mode
			\end{tabular}
		       &\begin{tabular}{p{4cm}}
		        \citet{sekiya83}
			\end{tabular}\\ 
\tableline
$\rho^\ast_{\rm Roche}$   & $3.5\frac{M_\ast}{r^3}$
                       & \begin{tabular}{p{5cm}} 
                          Required by satellite to resist tidal disruption by primary
			 \end{tabular}
		       & \begin{tabular}{p{4cm}}
		         \citet{chandra87} 
			 \end{tabular}\\ 
\tableline
$\rhotwo$ & $\frac{1}{2\pi} \frac{\Qg}{\Qd^{\ast 2}} \left( \frac{\Sigmag}{\Sigmad} \right)^2 \frac{\Hdagger}{\Hg} \frac{M_\ast}{r^3} \sim 2 \times 10^4 \frac{M_\ast}{r^3}$  \tablenotemark{(b)}
                       & \begin{tabular}{p{5cm}} 
		         Equivalent to $\Qd<\Qd^\ast$ for dust-rich sublayers in gas
			 \end{tabular}
                       & \begin{tabular}{p{4cm}}
		         This paper, equation (\ref{eqn:rho00})
			 \end{tabular}\\

\enddata \label{tab:tab1}
\tablenotetext{(a)}{~Value is derived for $\Qg^\ast=1$ and $\Hg^\dagger/\Hg = 1$.}
\tablenotetext{(b)}{~Value is derived for $\Qd^\ast=1$, $\Qg=30$, $\metal=0.015$, and $\Hg^\dagger/\Hg = 1$.}
\end{deluxetable}

In \S\ref{sec:method}--\S\ref{sec:result}, we present
numerical simulations of 3D, self-gravitating, compressible flows of thin,
dense sublayers of dust. We use these simulations to try to
identify which of the proposed criteria (if any) is the most
relevant for gravitational instability. Section \ref{sec:discuss} summarizes our findings but
also points out the limitations of our numerical simulations, which
are restricted to the asymptotic limit $\Omega t_{\rm stop}
\rightarrow 0$. We argue in \S\ref{sec:wayout} how finite but
still small values of $\Omega t_{\rm stop}$ may lower the threshold for gravitational collapse.

\subsection{Critical Density for Gravitational \\ Instability in a Pure Gas Disk} \label{sec:intro_gas}

The usual criterion for gravitational instability in a razor-thin
pure gas disk is expressed in terms of the dimensionless parameter
\begin{equation} \label{eqn:Q_g}
Q_{\rm g} \equiv \frac{\cg \Omega}{\pi G \Sigmag}
\end{equation}
where $G$ is the gravitational constant, $\cg$ is the gas sound
speed, and $\Sigmag$ is the gas surface density (\citealt{goldreichbell65}; \citealt{toomre64}; \citealt{toomre81}). In (\ref{eqn:Q_g}), the Kepler
orbital frequency $\Omega$ has been substituted for the radial epicyclic
frequency. If
\begin{equation} \label{eqn:Q_g_crit}
\Qg < \Qg^\ast = 1 \,,
\end{equation}
the disk is gravitationally unstable to axisymmetric perturbations in
the disk plane. The $Q$-criterion is a measure
of the competition between stabilizing pressure, stabilizing rotation,
and de-stabilizing self-gravity (see, e.g., \citealt{binneytremaine}).
When $\Qg > 1$, horizontal perturbations having lengthscales $< 2\cg/G\Sigmag$
are stabilized by pressure, while those having lengthscales $> 2\cg/G\Sigmag$
are stabilized by rotation. When $\Qg$ equals 1, the first axisymmetric mode to become
unstable to self-gravity has radial wavelength $2\cg/G\Sigmag$. And as $\Qg$ approaches $1$ from above, the disk is
increasingly susceptible to nonaxisymmetric perturbations which swing
amplify (\citealt{goldreichbell65}).

The criterion $\Qg \lesssim \Qg^\ast$ for gravitational instability
can be translated into a criterion for the midplane density
$\rhogi$ (the subscript ``0'' denotes the initial midplane value). We define
a disk half-thickness $\Hg$ using
\begin{equation} \label{eqn:half_a}
\Sigmag \equiv 2 \rhogi \Hg \,.
\end{equation}
We also define a half-thickness $\Hg^\dagger$ using the usual relation
from vertical hydrostatic equilibrium:
\begin{equation} \label{eqn:half_b}
\Hg^\dagger \equiv \cg/\Omega \,.
\end{equation}
Ordinarily $\Hg \approx \Hg^\dagger$ and we would not bother to distinguish the two; however, we will later find cases where they differ by factors of several
because of the effects of dust, and thus we take care to separate the two lengths now. Upon substitution of (\ref{eqn:half_a}) and (\ref{eqn:half_b}), the relation $\Qg \lesssim \Qg^\ast$ is shown to be equivalent to
\begin{equation} \label{eqn:rhoI}
\rhogi \gtrsim \rhoone = \frac{1}{2\pi} \frac{1}{\Qg^\ast} \frac{\Hg^\dagger}{\Hg} \rho^\dagger
\end{equation}
where we have defined a reference density\footnote{In this paper, we will superscript critical threshold densities with $\ast$, and fiducial or reference quantities with $\dagger$.}
\begin{equation} \label{eqn:dagger}
\rho^\dagger \equiv M_\ast / r^3
\end{equation}
with $M_\ast$ and $r$ equal to the mass of the central star and the disk radius,
respectively. 

The $\rhoone$-criterion (\ref{eqn:rhoI}) is sometimes used (e.g., \citealt{leeetal10,leeetal10b}) to signal gravitational
instability in dusty gas disks (with $\rhogi$ replaced by the total
dust + gas density $\rhodi+\rhogi$, $\Qg^\ast = 1$, and $\Hg^\dagger/\Hg = 1$). 
But using $\rhoone$ for dust-gas mixtures is suspect
because the criterion does not account explicitly for the two-phase nature of
such media. In the next subsection we make such an accounting
to derive a substantially different criterion for gravitational collapse.


\subsection{Critical Density for Gravitational \\ Instability in a Dust-Rich Sublayer \\ in the Limit $\Omega t_{\rm stop} \rightarrow 0$} \label{sec:intro_mixed}

For disks of gas and dust, gravitational instability should still be
determined by the $Q$-criterion, except there is now the possibility
that disk self-gravity is dominated by dust in a vertically thin sublayer
at the midplane:
\beq \label{eqn:Q_d} \Qd \equiv \frac{\cd \Omega}{\pi G
  \Sigmad} \lesssim \Qd^\ast \,\,\, {\rm for \,\, instability.}
\enq 
In using the
dust surface density $\Sigmad$ in (\ref{eqn:Q_d}), we neglect the
contribution of gas to the total surface density of the sublayer.
Under typical circumstances, the error accrued is small.

In the limit $\Omega t_{\rm stop} \rightarrow 0$, the dust-gas mixture
represents a colloidal suspension. In this suspension, dust does not
contribute to the pressure $P$ --- which is still provided entirely by
gas --- but instead adds to the inertia. In other words,
\beq \label{eqn:eos} P = \rhog \cg^2 = (\rhog + \rhod) \cd^2 \enq by
definition of $\cd$, the speed of sound in the suspension:
\beq \label{eqn:c_d} \cd = \frac{\cg}{\sqrt{1+\mu}} \enq 
where $\rhog$ is the local gas density, $\rhod$ is the local dust density, 
and $\mu = \rhod/\rhog$ is the dust-to-gas ratio. In effect, dust increases
the mean molecular weight of the gas.

Inserting (\ref{eqn:c_d}) into (\ref{eqn:Q_d}) and using
\begin{equation}
\rhogi = \rho^\dagger \frac{1}{2\pi \Qg} \frac{\Hg^\dagger}{\Hg} \,,
\end{equation}
we solve for the total
midplane density required for gravitational instability:
\begin{equation} 
\rho_0 = \rhodi + \rhogi \gtrsim \rhotwo 
\end{equation}
where
\begin{align}
\rhotwo & = \frac{1}{2\pi}\frac{\Qg}{\Qd^{\ast 2}} \left( \frac{\Sigmag}{\Sigmad} \right)^2 \frac{\Hg^\dagger}{\Hg} \, \rho^\dagger \label{eqn:rho00}\\
& \approx  2 \times 10^4 \rho^\dagger
\left( \frac{\Qg}{30} \right) \left( \frac{1}{Q_d^\ast} \right)^2 \nonumber\\
& \left( \frac{0.015}{\Sigmad/\Sigmag} \right)^2 \left( \frac{\Hg^\dagger/\Hg}{1} \right) \,.\label{eqn:rho0}
\end{align}
In (\ref{eqn:rho0}), our normalizations for $\Qg$ and the bulk (height-integrated but local to $r$) metallicity
$\Sigmad/\Sigmag$ derive from the MMSN at $r = 1$ AU
\citep{chiangyoudin10}. For these parameter choices, the critical midplane
density $\rhotwo$ is an astonishing five orders of magnitude
greater than $\rhoone$. It is possible that real disks have masses
and bulk metallicities enhanced over the MMSN by factors of a few, in which
case $\rhotwo$ would be larger than $\rhoone$ by about three orders of magnitude.

\subsection{Other Critical Densities} \label{sec:sekiya}

Another threshold density, already alluded to at the beginning of \S\ref{sec:intr}, is the Roche density:
\begin{equation}
\rho^\ast_{\rm Roche} = 3.5 \frac{M_\ast}{r^3} \,.
\end{equation}
The Roche density is the density required for a strengthless,
incompressible, fluid body in hydrostatic equilibrium to resist tidal
disruption, when in synchronous orbit at distance $r$ about a star 
of mass $M_\ast$ \citep[e.g.,][]{chandra87}.

Yet another candidate threshold was proposed by \citet{sekiya83}, who found that when the midplane density exceeds
\begin{equation}
\rho^\ast_{\rm Sekiya} = 0.60 \frac{M_\ast}{r^3} \,,
\end{equation}
the disk becomes susceptible to an unstable, incompressible,
axisymmetric mode in which in-plane motions generate out-of-plane
bulges (i.e., an annulus that contracts radially becomes thicker
vertically, and vice versa). The nonlinear outcome of this instability
is not known, but \citet{sekiya83} speculated that the dust sublayer
might eventually fragment on the scale of the wavelength of the
overstable mode, and that dust particles might sediment toward the centers
of fragments to form the first-generation planetesimals.

Table \ref{tab:tab1} summarizes the four candidate threshold densities.
For realistic parameters ($\Qg \sim 10$--30; $\Sigmad/\Sigmag \sim 0.015$--0.15), the four densities obey
\begin{equation}
\rhoone < \rho^\ast_{\rm Sekiya} < \rho^\ast_{\rm Roche} \ll \rhotwo \,.
\end{equation}
The smallest three densities in this hierarchy are fixed multiples of the
reference density $\rho^\dagger = M_\ast / r^3$ (with coefficients $\sim$$1/2\pi \approx 0.16$, 0.6, and 3.5, respectively). The last density $\rhotwo$ can, in principle, be arbitrarily larger than $\rho^\dagger$; for typical, astrophysically plausible parameters, it is 2--4 orders of magnitude larger.

Which of the four densities in Table \ref{tab:tab1} is the most
accurate predictor of gravitational collapse? In the next two
sections, we describe numerical simulations performed in the $\Omega
t_{\rm stop} \rightarrow 0$ limit that attempt to answer this
question. We will find unfortunately that the numerical expense of simulating
thin sublayers of dusty gas will force us into a parameter space
where the difference between $\rhotwo$ and
the other densities is not as large as it is in reality;
we will have to make do with what we can.


\section{METHODS}
\label{sec:method}

\subsection{Code \label{sec:numerical}}

We simulate hydrodynamic, self-gravitating, stratified
flows in disks using \texttt{Athena},
configured for a shearing box, with no magnetic fields
\citep{stoneetal08,sg10}.  Dust is assumed to be perfectly
aerodynamically coupled to gas so that they share the
same velocity field $\mathbf{v}$.

The equations solved are:
\begin{align}
\frac{\partial \rho}{\partial t} + \nabla\cdot (\rho \mathbf{v}) = 0 \,, 
\label{eq:continuity1} \\
\frac{\partial \rhog}{\partial t} + \nabla\cdot (\rhog \mathbf{v})= 0 \,, 
\label{eq:continuity2} \\
\frac{\partial \rho\mathbf{v}}{\partial t} + \nabla\cdot\left(\rho\mathbf{v}\mathbf{v}
  +\mathbf{P}\right) = - \rho\nabla\Phi \nonumber \\
- 2\rho(\Omega \hat{\mathbf{z}})\times\mathbf{v} +2q\rho\Omega^2 x \hat{\mathbf{x}}
-\rho\Omega^2 z\hat{\mathbf{z}}\,,
\label{eq:eom_hill} \\
\nabla^2\Phi = 4\pi G \rho\,,
\end{align}
where $\rho=\rhog+\rhod$ is the total density of the dust-gas suspension, $\mathbf{P}=P \mathbf{I}$
is a diagonal tensor with components $P=\rhog \cg^2$ as defined in
equation~(\ref{eqn:eos}) with constant $\cg$ (isothermal approximation), $\Omega$ is the mean (constant) orbital frequency, $\hat{\mathbf{x}}$ points in the radial direction, $\hat{\mathbf{z}}$ points in the vertical direction, and
$\Phi$ is the self-gravitational potential of the dust-gas mixture. We choose the shear parameter
$q=3/2$ for Keplerian flow.

\subsubsection{Algorithms and boundary conditions}\label{sec:abc}

\texttt{Athena 4.0} provides several schemes for
time integration and spatial reconstruction, and for solving the
Riemann problem. Having experimented with various options, we adopted
the van Leer algorithm for our dimensionally unsplit integrator
\citep{vl06,sg09}; a piecewise linear spatial reconstruction in the
primitive variables; and the HLLC (Harten-Lax-van Leer-Contact)
Riemann solver. To account for disk self-gravity,
we use the routines written by \citet{ko09} and \citet{kko11}
which solve Poisson's equation using fast Fourier transforms.

Boundary conditions for our hydrodynamic flow variables (including
density and velocity, but not the self-gravitational potential) are
shearing-periodic in radius ($x$) and periodic in azimuth
($y$).  For vertical height ($z$), we experimented with both periodic
and outflow boundary conditions, and chose periodic boundary
conditions to ensure strict mass conservation. When outflow boundary conditions
were employed, mass was lost from the boundaries at early times 
and complicated the interpretation of our results. We verified
that our results are insensitive to box height for sufficiently tall
boxes; see \S\ref{sec:result} for explicit tests.


The Poisson solver implements shear-periodic boundary conditions in $x$,
periodic boundary conditions in $y$, and vacuum boundary conditions in $z$
\citep{ko09,kko11}. In our simulations, self-gravity is dominated by dust,
and our boxes are tall enough to contain the entire dust layer.
Both vertical and radial stellar tidal gravity are included as source terms
in the van Leer integrator.

We further augmented the code to include dust in the limit of zero
stopping time. In this limit, dust shares the same velocity field as
gas, and contributes only to the mass density.
In our modified version of \texttt{Athena}, two continuity equations
are solved: one for the entire mixture ($\rho = \rhog + \rhod$, see
equation~\ref{eq:continuity1}), and
one for the gas ($\rhog$, see equation~\ref{eq:continuity2}).  The dust density is given by the
difference ($\rho - \rhog$). The remaining hydrodynamic equations
govern the dust-gas mixture ($\rho$), but with gas ($\rhog$)
contributing solely to the pressure $P$ (see equation~\ref{eq:eom_hill}).  For simplicity, we adopt an
isothermal equation of state so that $P$ is related to $\rhog$ by
equation~(\ref{eqn:eos}) for constant $\cg$. Isothermal flows are more
prone to gravitational instability than adiabatic ones \citep{mamarice10}.

We also modified the HLLC Riemann solver to accommodate our dust-gas
mixture.  Changes include the following: (1) The speeds of the left,
right, and contact waves are reduced by a factor $(1+\mu)^{-1/2}$,
where $\mu \equiv \rhod/\rhog$ is the local dust-to-gas ratio, to
account for the added inertia from dust (see equation~\ref{eqn:c_d}).
(2) The pressure in the contact region is replaced by an equivalent
but numerically more accurate form based on equation~(10.76) in
\citet{toro99}.
(3) When calculating left/right momentum fluxes, we ensure that
only gas contributes to the pressure by using $\rhog$ and not $\rho$.
(4) For the flux solver to predict the pressure and wave
speeds, the left/right gas densities require specification. We
therefore add a reconstruction process for the gas density which
interpolates cell-centered values to cell boundaries to second-order
accuracy.

Previous studies of dust in the perfectly coupled limit \citep{chiang08,leeetal10,leeetal10b}
also introduced a static background radial pressure gradient to mimic
sub-Keplerian rotation of gas in a pressure-supported disk.  We could
also add the appropriate source term to the van Leer integrator.
However, since our goal is to determine the minimum densities required
for gravitational collapse and not to study vertical shear
instabilities (i.e., the Kelvin-Helmholtz instability), we omit the
background pressure gradient for simplicity.

In many of our simulations, the dust layer at the midplane collapses
vertically because it is gravitationally unstable. Because of our
boundary conditions, ``fresh'' gas from outside the simulation box
cannot enter into the box, and thus in the event of gravitational
collapse toward the midplane, the topmost and bottommost regions of
our simulation domain become evacuated. Low-density gas in those
regions become increasingly easy to accelerate, and the code timestep
shortens by orders of magnitude, effectively halting the simulation.
The dramatic reduction in timestep is not a serious limitation, as it
usually occurs after the collapsing dust has attained some saturated
state (see \S\ref{sec:standard_run}).  In any case, we are more
interested in the onset of gravitational instability than its
nonlinear development.

\subsubsection{Code tests}
The following test problems helped to validate our code.

\textsl{Linear wave propagation.---} We propagated a small-amplitude 1D wave
in a medium with a uniform background dust-to-gas ratio, with periodic boundary conditions, no background shear, and no gravity.  We found the simulated wave speed matched the reduced sound speed calculated in (\ref{eqn:c_d}).
We chose our box to be one wavelength long, so that after one
wave period, the wave crossed the boundaries and returned to its
original position. With $N = 128$ grid cells and an initial
(fractional) wave amplitude $A=10^{-4}$, we found the deviation $\delta
q \equiv \frac{1}{N}\displaystyle{\sum^{N}_{i=1}}|q_i-q_i^0| \approx 
2\times 10^{-8}$, where $q \in \{\rhod,\rhog,\rho\}$ and $q^0$ represents
the initial condition.

\textsl{Dust cloud advection.---} We advected a Gaussian-shaped dust
cloud in a 1D domain.  The cloud occupied about half the size of the
box and the code was run for one box-crossing time. With $N=256$ grid
cells, the root-mean-squared deviation in the shape of the cloud was $<
1$\%.

\textsl{Hydrostatic equilibrium of a stratified but
  non-self-gravitating dusty disk.---} Omitting self-gravity but
including stellar gravity (both radial and vertical), we set up 3D
dust-gas mixtures in hydrostatic equilibrium. A variety of vertical
profiles for the dust-to-gas ratio were tested, ranging from uniform
to linear to more complicated functional forms. All equilibria were
found to be stable against small perturbations, even for dust-to-gas
ratios as large as several hundred.

\textsl{Gravitational instability of 3D pure gas disks.---} We
simulated isothermal, gravitationally unstable disks of pure gas in
3D. The gas was initialized in hydrostatic equilibrium (computed with
vertical self-gravity), and box heights spanned approximately $\pm4$
initial gas scale heights. We found that $\Qg = 1$ did not trigger
gravitational instability, whereas $\Qg = 0.5$ did. Our results are
consistent with those of \citet{goldreichbell65}, who found
analytically that $\Qg^\ast = 0.676$ for a finite-thickness isothermal
gas disk.



\subsection{Initial Conditions \\ and Run Parameters \label{sec:initial}}

Initial conditions for our science simulations are of a dust-gas mixture with a pre-defined
vertical profile for the dust-to-gas ratio $\mu(z) = \rhod(z)/\rhog(z)$.
We choose the form
\beq
\mu(z) \equiv \mu_0 \, \sech^2\left(\frac{z}{\zd}\right), 
\label{eq:mu}
\enq
where $\mu_0$ is the midplane dust-to-gas ratio. The scale height
$\zd$ can be thought of as the half-thickness
of the dust layer insofar as $\rhog(z)$ is constant
with $z$.

The isothermal dust-gas mixture is initialized in vertical hydrostatic equilibrium, including both stellar tidal gravity and disk self-gravity:
\beq
\frac{\cg^2}{\rhog+\rhod}\frac{d\rhog}{d z} = -\Omega^2 z - 4\pi
G\int^z_0(\rhog+\rhod)dz .
\label{eq:static1}
\enq
We solve numerically the differential form of (\ref{eq:static1}).
Taking derivatives, we find
\beq
\frac{d}{dz}\left[(1+\mu)^{-1}\frac{d\ln \rhog}{dz}\right] = -\frac{1}{\Hdagger} - \frac{4\pi
G}{\cg^2}\rhog(1+\mu),
\label{eq:static2}
\enq where $\Hdagger \equiv \cg/\Omega$ is a fiducial (constant) gas scale height, not to be confused with any actual disk scale height.
A non-dimensional form of (\ref{eq:static2}) is given by
\beq
\frac{d}{d\tilde z}\left[(1+\mu)^{-1}\frac{d\ln\tilde{\rho}_{\rm g}}{d\tilde z}\right ] = -1
-\frac{2}{h_{\rm g}\Qg} \tilde{\rho}_{\rm g}(1+\mu), 
\label{eq:static3}
\enq
where we have defined the dimensionless variables $\tilde z\equiv
z/\Hdagger$, $\tilde{\rho}_{\rm g}\equiv \rhog/\rhogi$ (where $\rhogi$ is
the midplane gas density), and $h_{\rm g}\equiv\Hg/\Hdagger$, with
\begin{equation}
\Hg \equiv \Sigmag / (2 \rhogi) \,.
\end{equation}

Upon insertion of (\ref{eq:mu}), equation~(\ref{eq:static3}) can be
solved numerically for $\tilde{\rho}_{\rm g} (\tilde z)$. But the solution
must satisfy the following two constraints: 
\beq
h_{\rm g} = \int^{\infty}_{0}\tilde{\rho}_{\rm g}(\tilde{z}) \,d\tilde{z}
\label{eq:hg}
\enq
by definition of $\Hg$, and
\beq
\frac{\Sigma_{\rm d}}{\Sigma_{\rm g}} =
\frac{1}{h_{\rm g}}{\int^{\infty}_{0}\tilde{\rhog}(\tilde{z})\mu(\tilde{z})d\tilde{z}}
\label{eq:metal}
\enq
for a fixed height-integrated (i.e., bulk) metallicity $\metal$.

Our procedure is as follows. We freely specify $\Qg$, $\metal$, and
$\mu_0$ as model input parameters. We then iteratively solve
equations~(\ref{eq:static3}), (\ref{eq:hg}) and (\ref{eq:metal}) for
the three unknowns $\tilde{\rho}_{\rm g}(\tilde{z})$, $h_{\rm g}$,
and $\zd$. First we guess $\zd$ and $h_{\rm g}$, and integrate
(\ref{eq:static3}) to obtain $\tilde{\rho}_{\rm g} (\tilde z)$. If
$\tilde{\rho}_{\rm g}$ so calculated fails (\ref{eq:hg}), then we
revise $h_{\rm g}$ and re-integrate (\ref{eq:static3}), repeating until
(\ref{eq:hg}) is satisfied. Next we check (\ref{eq:metal}). If
$\tilde{\rho}_{\rm g}(\tilde z)$ and $h_{\rm g}$ fail
(\ref{eq:metal}), then we revise $\zd$ and repeat the procedure from
the beginning, re-integrating (\ref{eq:static3}) to obtain
$\tilde{\rho}_{\rm g}$, re-establishing (\ref{eq:hg}), and so
on. Typically $\sim$100 iterations ($\sim$10 for $\zd$ $\times$ $\sim$10 for $h_{\rm g}$)
are required before all constraints are satisfied to $\sim$1\% accuracy in $\zd$
and $10^{-6}$ accuracy in $h_{\rm g}$.

Table~\ref{tab:tab3} lists the parameters of our models.  Note that these parameters
do not describe plausible protoplanetary gas
disks; in particular, our model metallicities $\metal$ are
orders of magnitude above the solar value of $\sim$0.015. Parameters
are instead chosen to yield disk flows that our code can adequately
resolve while still testing equation (\ref{eqn:rho0}). Unfortunately,
more astrophysically realistic parameters correspond to dust sublayers
that are too vertically thin for us to resolve numerically; the code
timestep, set by the sound-crossing time across a grid cell, becomes
prohibitively short as thinner dust layers are considered. This difficulty
means that the difference between $\rhotwo$ and the other
candidate threshold densities is much less than what it 
is in reality, and our ability to distinguish between the candidates
degrades as a result.

\begin{figure}[h!]
\epsscale{1.05}
\plotone{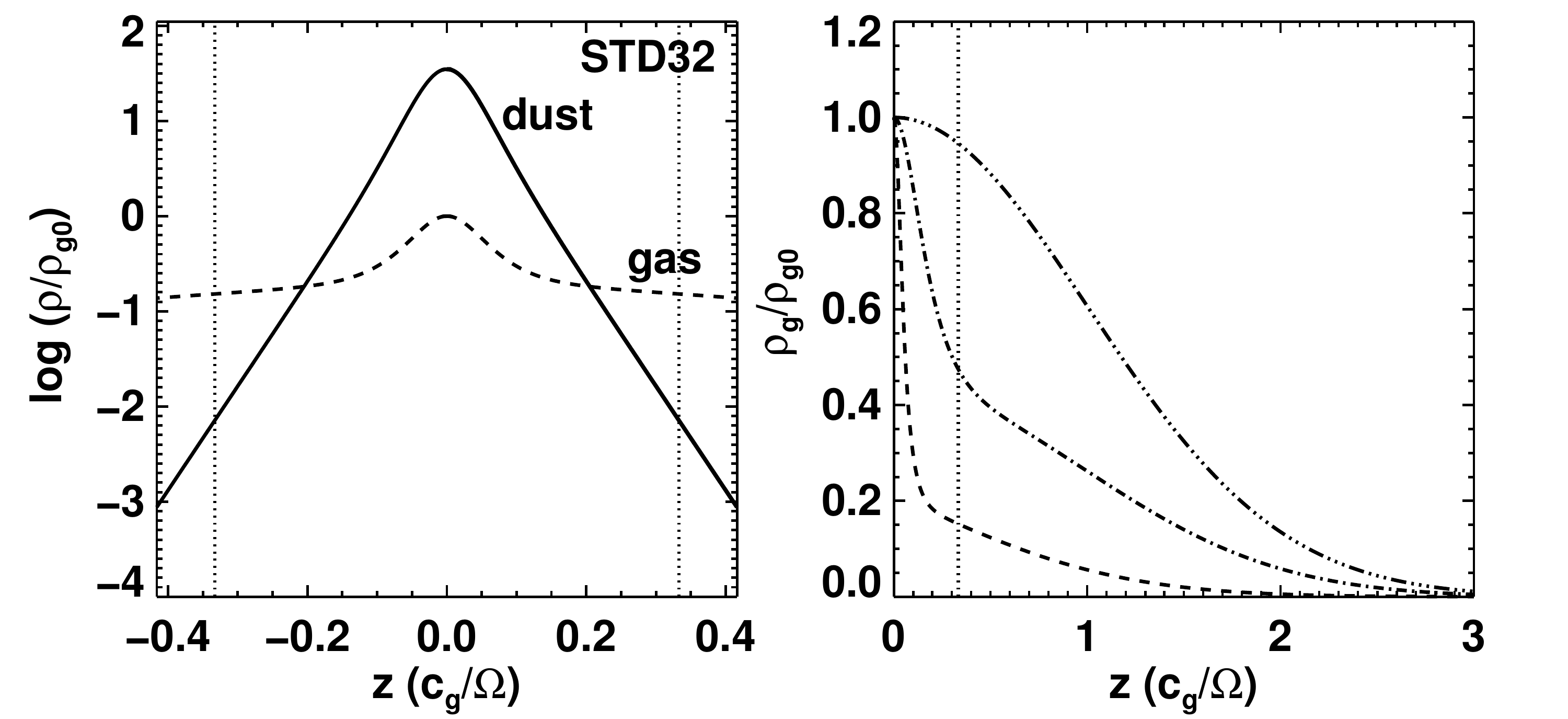} \caption{\small{Left: Initial dust and gas densities for standard run S = STD32. Right: Initial gas density (dashed) on an expanded scale, together with the gas density computed without self-gravity but with dust (dash-dot) and without dust but with self-gravity (dash-double-dot). Vertical lines in both panels lines delimit the top and bottom of our computational box.}}\label{fig:init0}
\end{figure}

Figure~\ref{fig:init0} plots the initial conditions for our standard
model (S = STD32), for which $\Qg=24$, $\metal=8$, and $\mu_0=35$.  For this
specific case, we calculate that $h_{\rm g}=0.20 \, \cg/\Omega$ and
$\zd=0.083 \,\cg/\Omega$.  The top and bottom boundaries of our
simulation box are indicated by dotted vertical lines; typically box
heights span $\pm$$4\zd$ (see \S\ref{sec:result} for box height
tests). The right-hand panel of Figure \ref{fig:init0} compares gas
density profiles computed with and without self-gravity, and with and without dust,
and shows that both the weight and self-gravity of the embedded dust
layer force gas into a similarly thin layer.

\begin{figure}[h!]
\epsscale{1.0}
\plotone{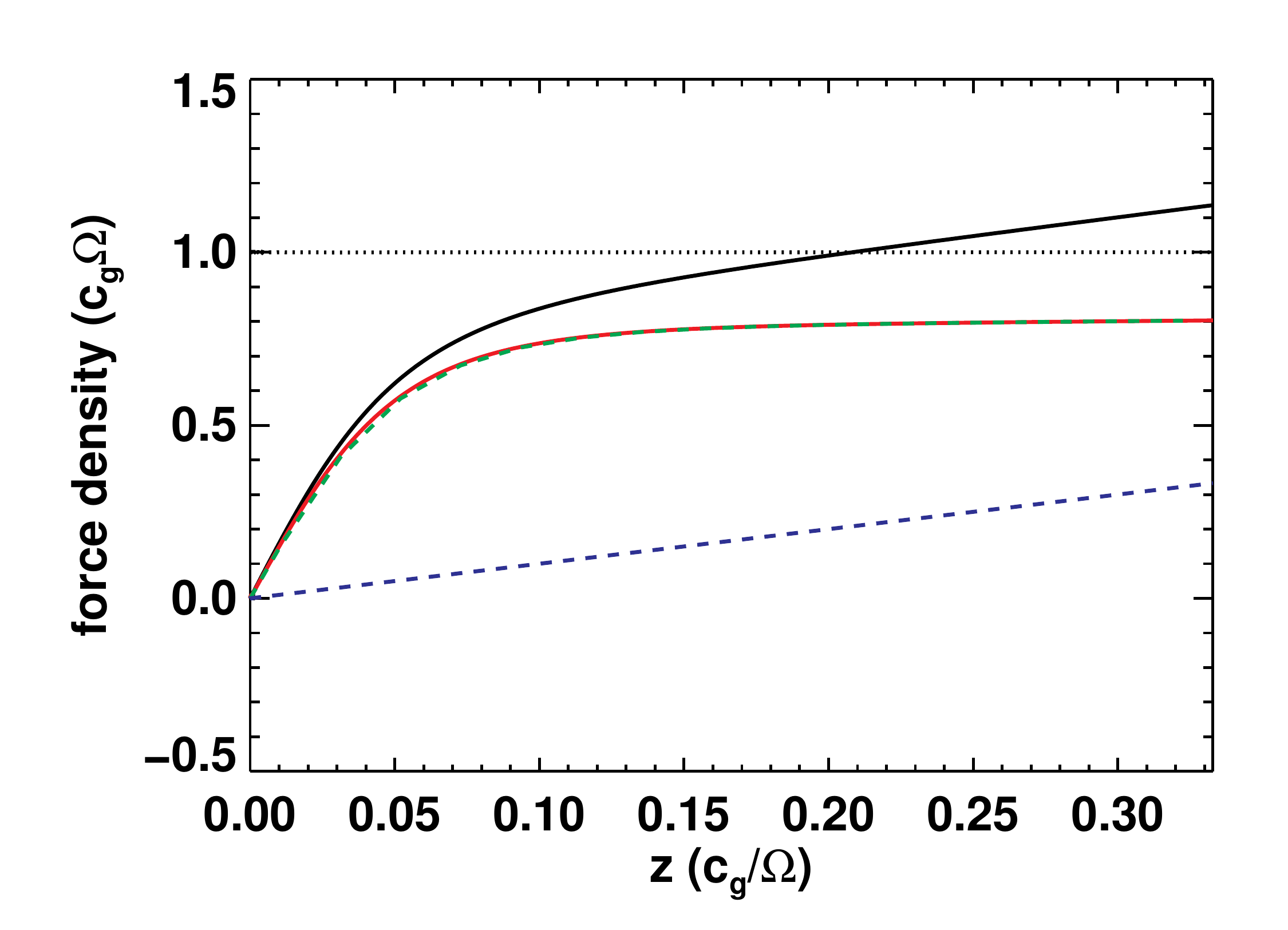} \caption{\small{Vertical force balance for our initial conditions. The force density due to self-gravity is computed two ways: by direct integration of the density profile (red solid), and by using the code's Poisson solver (green dashed). The two methods agree. The force density due to the pressure gradient (black solid) should equal the sum of self-gravity and the static stellar potential (blue dashed). The horizontal gray dotted line shows the ratio of
pressure to gravity. All force densities are shown in their absolute values. }}\label{fig:init1}
\end{figure}


Figure~\ref{fig:init1} plots the initial force densities within the upper
half of the simulation box to demonstrate how well 
vertical hydrostatic equilibrium is satisfied.
The sum of stellar gravity (blue dashed curve) and
disk self-gravity (red solid curve computed via the integral
in equation \ref{eq:static1}) should equal
the pressure gradient (black solid curve). It does,
as evidenced by the ratio of pressure to gravity (gray dotted line) which is practically constant at unity.
We also overplot the self-gravitational force
computed by our 3D Poisson solver (green dashed curve); the agreement
with the exact solution is good.

Every simulation listed in Table~\ref{tab:tab3} is perturbed from its initial
equilibrium by adding random cell-to-cell fluctuations of amplitude
$\sim$$10^{-3} \cg$ to the velocity field. The typical duration
of a simulation is $\sim$20 $\Omega^{-1}$. Our rationales
for box size and resolution are explained in \S\ref{sec:result}.



\section{RESULTS \label{sec:result}}
Results for 2D shearing sheets are described in \S\ref{sec:ssheet},
and those for 3D shearing boxes are in \S\ref{sec:sublayer}.

\subsection{2D Shearing Sheet\label{sec:ssheet}}

\begin{deluxetable}{ccccccccc}
\tabletypesize{\footnotesize}
\tablecolumns{9} \tablewidth{0pc}
\tablecaption{2D Simulation Parameters}
\tablehead{\colhead{Name} & \colhead{$\Qg$} &
\colhead{$\mu_0$} & \colhead{$\Qd$} & \colhead{$\lambda_c$($\cg/\Omega$)} &
\colhead{$L_x\times L_y$ ($\cg^2/\Omega^2$)} & \colhead{Resolution} & \colhead{GI\tablenotemark{(a)} } &
\colhead{~Duration ($\Omega^{-1}$)}}
\startdata
S2D0 & $12$ & $8$   & $0.44$ & $0.93$ & $10\times10$   & $256\times256$ & Y & $5.9$  \\
S2D1 & $12$ & $8$   & $0.44$ & $0.93$ & $10\times10$   & $512\times512$ & Y & $6$    \\
S2D2 & $12$ & $8$   & $0.44$ & $0.93$ & $0.5\times0.5$ & $32\times32$   & N & $100$  \\
S2D3 & $12$ & $8$   & $0.44$ & $0.93$ & $0.5\times0.5$ & $128\times128$ & N & $100$  \\
S2D4 & $12$ & $4.2$ & $1.0$  & $2.8 $ & $30\times30$   & $256\times256$ & N & $100$  \\
S2D5 & $12$ & $2.3$ & $2.0$  & $6.9 $ & $70\times70$   & $256\times256$ & N & $100$  \\
S2D6 & $6 $ & $4.2$ & $0.5$  & $1.4 $ & $15\times15$   & $256\times256$ & Y & $6.2$  \\
\enddata \label{tab:tab2}
\tablenotetext{(a)}{~GI = Gravitational Instability. Y means $\max \Sigmad$ increases by orders of magnitude over a few dynamical times, and N means it does not.}
\end{deluxetable}

For two-dimensional dusty disks, the criterion for gravitational
instability reads
\beq \label{eqn:2ddusty}
\Qd=\frac{\cd\Omega}{ \pi G(\Sigmad+
  \Sigmag)} =\frac{\Qg}{(1+\mu_0)^{3/2}} < \Qdtwod^\ast \,.
\enq
We test this criterion by constructing a series of 2D shearing sheet
simulations with various values of $\Qg$ and $\mu_0$, thereby seeing if we
can converge on a unique value for
$\Qdtwod^\ast$.  Although total surface
densities can change during the simulation, the dust-to-gas
ratio stays fixed at its initial value because of our perfect-coupling
approximation. Initial conditions are as follows: for a given
domain size $L_x$ and $L_y$,
the flow velocity $\mathbf{v}=-\frac{3}{2}\Omega x \hat{e}_y$ and
the surface density $\Sigma=\Sigma_{0} +
\delta \Sigma \cos(\mathbf{k\cdot x})$, with $\Sigma_{0} = \Sigma_{\rm g0}
+\Sigma_{\rm d0} =\Sigma_{\rm g0}(1+\mu_0)$, $\delta\Sigma/\Sigma_0 = 0.01$,
$k_x = -2(2\pi/L_x)$, and $k_y = 2\pi/L_y$. In our 2D simulations,
we choose $\cg=\Omega=\Sigma_{\rm g0}=1$ as our units.

Table~\ref{tab:tab2} lists the parameters for our 2D runs.  Our
standard 2D run, labeled \textrm{S2D0}, has $\Qg=12$ and $\mu_0=8.0$
and therefore $\Qd=0.44$. For this run, the domain size is chosen
large enough to easily fit the critical wavelength $\lambda_c$ for
gravitational instability: $L_x=L_y=10\cg/\Omega\gtrsim 10\lambda_c$,
where
\beq \label{eq:critical_wavelength}
\lambda_c \equiv \frac{2\cd^2}{G\Sigma_0}
\enq
is the
wavelength of the fastest growing mode according to the WKB dispersion
relation for axisymmetric waves. It is also the wavelength of the
first mode to become unstable when $\Qd$ just crosses $\Qdtwod^\ast$.  The
resolution of the standard run is $N_x \times N_y = 256\times256$ so
that one critical wavelength is resolved across $\sim$$10$ grid cells.

\begin{figure}[h!]
\plotone{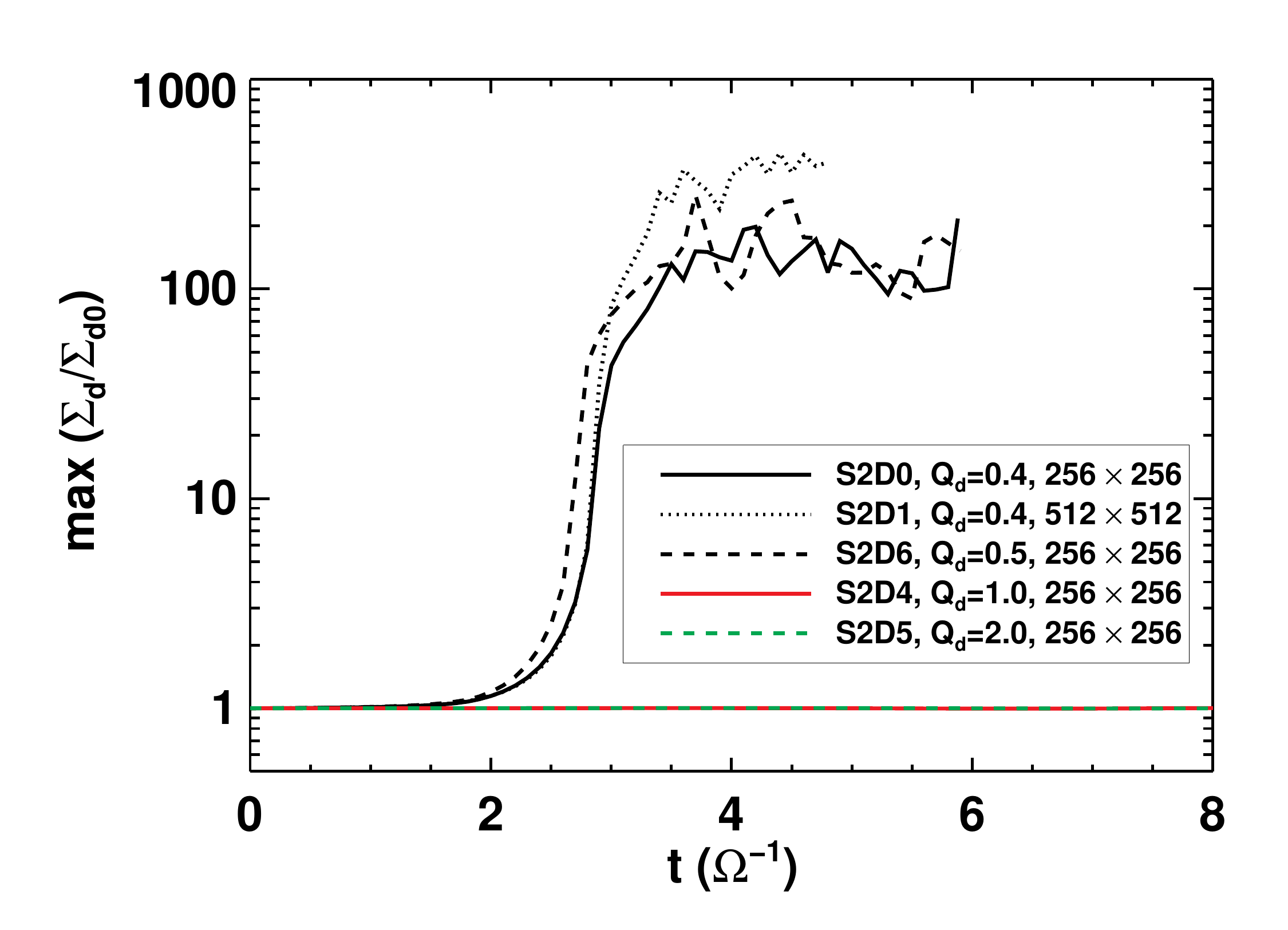} \caption{\small{Time evolution of the maximum dust density
$\max \,\Sigmad$ for our 2D shearing sheet simulations. The critical value
$Q_{\rm d,2D}^\ast$ below which gravitational instability is triggered appears to be between 0.5 and 1.0.}}\label{fig:dmax_2d}
\end{figure}

For \textrm{S2D0}, we find that the disk is indeed gravitationally
unstable: density waves steepen quickly, and dense clumps of dusty gas
form before one orbital period elapses. A simple way to
portray instability is to track the maximum dust density
$\max \Sigmad$ versus time --- this is done in Figure \ref{fig:dmax_2d},
which shows that the maximum dust density increases by two orders
of magnitude over a few dynamical times for our standard run.

Also shown in Figure \ref{fig:dmax_2d} are results for other runs.
In \textrm{S2D4}, \textrm{S2D5}, and \textrm{S2D6}, either
$\Qg$ or $\mu_0$ is varied relative to our standard run, so that
$\Qd$ varies from 0.5 to 2.0.
For all these runs, the domain size is $\sim$10$\lambda_c$ in each
direction and the resolution is $\sim 10$ cells per $\lambda_c$, just
as in the standard case. Taken together, the results indicate that 
\begin{equation}
0.5 < Q_{\rm d,2D}^\ast < 1.0 \,.
\end{equation}

Other runs explore the effects of
varying resolution and domain size. Doubling both $N_x$ and $N_y$
relative to our standard run (\textrm{S2D1})
enables higher maximum densities to be
achieved when the instability saturates, but otherwise does not seem
to alter the evolution. 
Reducing the size of the box so that it can no longer accommodate even a single critical wavelength (\textrm{S2D2}, \textrm{S2D3}) results in no instability, as expected \citep[][]{gammie01,JG2003}.

\subsection{3D Stratified Dusty Disks \label{sec:sublayer}}
Equation (\ref{eqn:Q_d}; equivalently \ref{eqn:2ddusty}) gives the criterion for gravitational
instability in a 2D razor-thin sheet. For a 3D, vertically stratified
disk, there is some ambiguity as to how we evaluate $\cd$ in equation
(\ref{eqn:Q_d}) because its value varies with height. Here we simply
take $\cd$ to be its value at the midplane, so that criterion
(\ref{eqn:Q_d}) becomes 
\beq \Qd \simeq \Qg \,\frac{1}{\metal} \,
\frac{1}{(1+\mu_0)^{1/2}} \lesssim \Qd^\ast \,.  \enq 
An alternative
is to calculate a vertically averaged, density-weighted sound speed.
We found, however, that such a procedure made little practical difference,
since dust densities are much greater than gas densities near the
midplane and drop steeply with height.

\subsubsection{Standard run (\textrm{S} = \textrm{STD32})\label{sec:standard_run}}
To orient the reader, we present results for
our standard 3D run (\textrm{S}, also labeled
\textrm{STD32} in \S\ref{sec:converge}), for which $\Qd = 0.5$.
The full set of model \textrm{S} parameters are listed in
Table~\ref{tab:tab3}, and the initial gas and dust density profiles
are displayed in Figure~\ref{fig:init0}.
Our simulation box extends $\pm$$4\zd$ vertically, and $14\zd$ in either
horizontal direction.
Each horizontal length is about twice the critical wavelength
($\lambda_c \approx 6.3 \zd$). The resolution is $32 \times 32 \times 32$
so that one horizontal critical wavelength spans $\sim$16 cells,
and one vertical scale length $\zd$ spans 4 cells.
These choices for domain size and resolution are tested
in \S\ref{sec:converge}. The simulation is terminated 
at $\sim$$10.3\Omega^{-1}$, at which point the timestep
has become three orders of magnitude smaller than the initial
timestep (see the final paragraph of \S\ref{sec:abc}).

{\renewcommand{\arraystretch}{1.2}
\begin{deluxetable}{ccccccccccccccc}
\tabletypesize{\scriptsize}
\tablecolumns{15} 
\tablewidth{0pt}
\tablecaption{3D Simulation Parameters (``Science Runs'')}
\setlength{\tabcolsep}{0.03in}
\tablehead{
\colhead{\begin{tabular}{c} Name        \\    \end{tabular} } &
\colhead{\begin{tabular}{c} $\Qg$       \\    \end{tabular} } &
\colhead{\begin{tabular}{c} $\mu_0$     \\    \end{tabular} } &
\colhead{\begin{tabular}{c} $\metal$    \\    \end{tabular} } &
\colhead{\begin{tabular}{c} $\Hg$       \\  ($\cg/\Omega$)  \end{tabular} } &
\colhead{\begin{tabular}{c} $\zd$       \\  ($\cg/\Omega$)  \end{tabular} } &
\colhead{\begin{tabular}{c} $\lambda_c$ \\  ($\zd$)  \end{tabular} } &
\colhead{\begin{tabular}{c} $L_x\times L_y\times L_z$   
                                        \\  ($\zd^3$)  \end{tabular} } &
\colhead{\begin{tabular}{c} Resolution  \\    \end{tabular} } &
\colhead{\begin{tabular}{c} Duration    \\  ($\Omega^{-1}$)  \end{tabular} } &
\colhead{\begin{tabular}{c} $\Qd$       \\    \end{tabular} } &
\colhead{\begin{tabular}{c} $\rho_0$    \\  ($\rho^\dagger$)  \end{tabular} } &
\colhead{\begin{tabular}{c} $\rhoone $\tablenotemark{(a)}    \\  ($\rho^\dagger$)  \end{tabular} } &
\colhead{\begin{tabular}{c} $\rhotwo $\tablenotemark{(b)}    \\  ($\rho^\dagger$)  \end{tabular} } &
\colhead{\begin{tabular}{c} GI\tablenotemark{(c)}          \\    \end{tabular} } 
}
\startdata
\textrm{S} & $24$ & $35.0$  & $8.0$  & $0.20$  & $0.083$ & $6.3$  &
$14\times14\times8$ & $32\times32\times32$ & $10.3$ & $0.5$ & $1.20$ & $0.16$ & $0.30$ & Y \\
\textrm{R1} & $24$ & $165.0$ & $2.0$ & $0.91$  & $0.011$ & $42.6$ &
$90\times90\times8$   & $256\times256\times32$ & $11$ &$0.93$ & $1.21$ & $0.16$ & $1.05$ & Y/N\tablenotemark{(d)}\\
\textrm{R2} & $12$ & $93.0$ & $0.67$ & $1.04$  & $0.008$ & $150.3$ &
$256\times256\times8$   & $256\times256\times32$ & $30$ & $1.86$ & $1.20$ & $0.16$ & $4.09$ & N\\
\textrm{R3} & $12$ & $143.0$ & $0.54$ & $1.07$  & $0.004$ & $255.4$ &
$400\times400\times8$   & $400\times400\times32$ & $30$ & $1.86$ & $1.77$ & $0.16$ & $6.12$ & N\\
\textrm{R4} & $12$ & $322.0$ & $0.56$ & $1.07$  & $0.002$ & $220.7$ &
$400\times400\times8$   & $400\times400\times32$ & $30$ & $1.2$ & $4.0$ & $0.16$ & $5.16$ & N\\
\textrm{R5} & $12$ & $322.0$ & $1.33$ & $0.87$  & $0.004$ & $44.7$ &
$400\times400\times8$   & $400\times400\times32$ & $3.6$ & $0.5$ & $4.9$ & $0.16$ & $1.24$ & Y\\
\textrm{SR} & $24$ & $35.0$ & $8.0$ &  $0.20$  & $0.083$ & $6.3$ &
$400\times400\times8$   & $400\times400\times32$ & $10.0$ & $0.5$ & $1.20$ & $0.16$ & $0.30$ & Y\\
\textrm{Z} & $24$ & $35.0$  & $4.0$  & $0.52$  & $0.077$ & $13.1$ & 
$30\times30\times8$ & $32\times32\times32$ & $30$   & $1.0$ & $0.46$ & $0.16$ & $0.46$ & N\\
\textrm{Q} & $48$ & $35.0$  & $8.0$ &  $0.35$  & $0.12 $ & $8.6 $ &
$20\times20\times8$ & $32\times32\times32$ & $30$   & $1.0$ & $0.34$ & $0.16$ & $0.34$ & N\\
\textrm{M} & $24$ & $8.0$   & $8.0$ &  $0.24$  & $0.71 $ & $3.2 $ &
$8\times8\times8$   & $32\times32\times32$ & $30$   & $1.0$ & $0.25$ & $0.16$ & $0.25$ & N\\
\enddata \label{tab:tab3}
\tablenotetext{(a)}{~Values are derived using $\Qg^\ast=1$.}
\tablenotetext{(b)}{~Values are derived using $\Qd^\ast=1$.}
\tablenotetext{(c)}{~GI = Gravitational Instability. Y means $\max \rhod$ increases by orders of magnitude over a few dynamical times, and N means it does not.}
\tablenotetext{(d)}{~See Figure \ref{fig:dmax_3d}.}
\end{deluxetable}
}

\begin{figure*}
\centering{
\includegraphics[width=8cm]{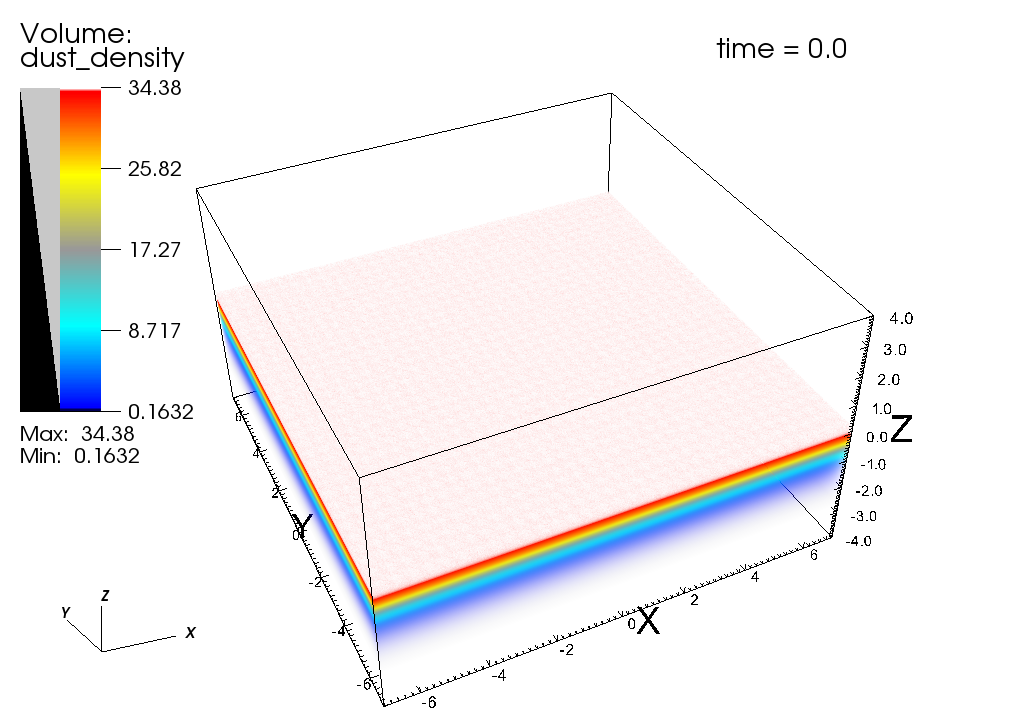}
\includegraphics[width=8cm]{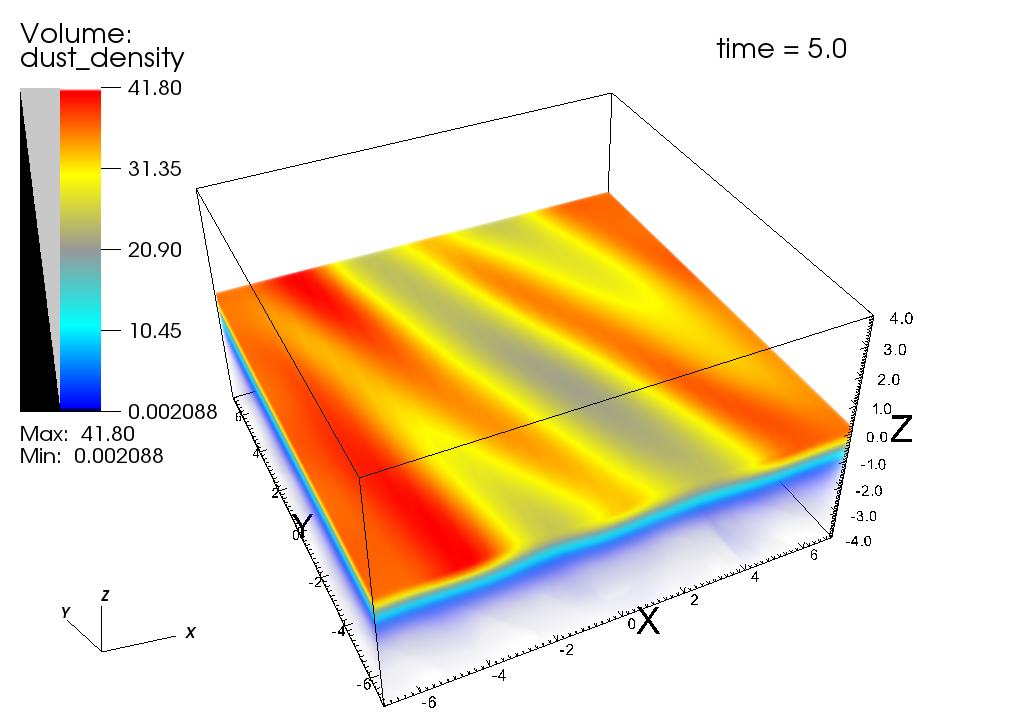} 
\includegraphics[width=8cm]{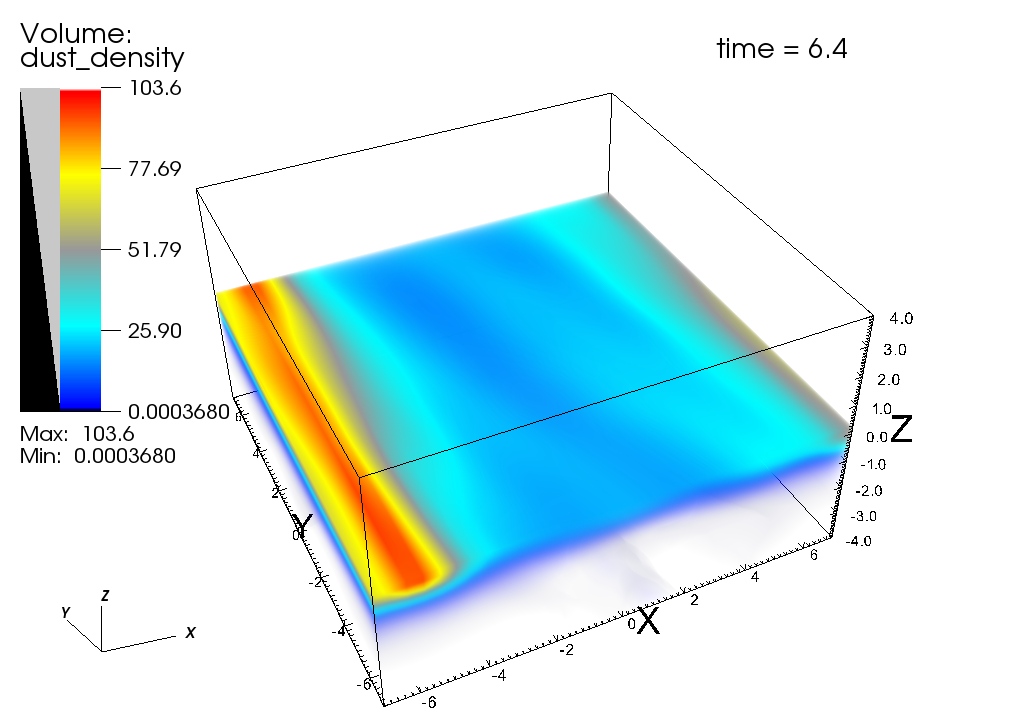} 
\includegraphics[width=8cm]{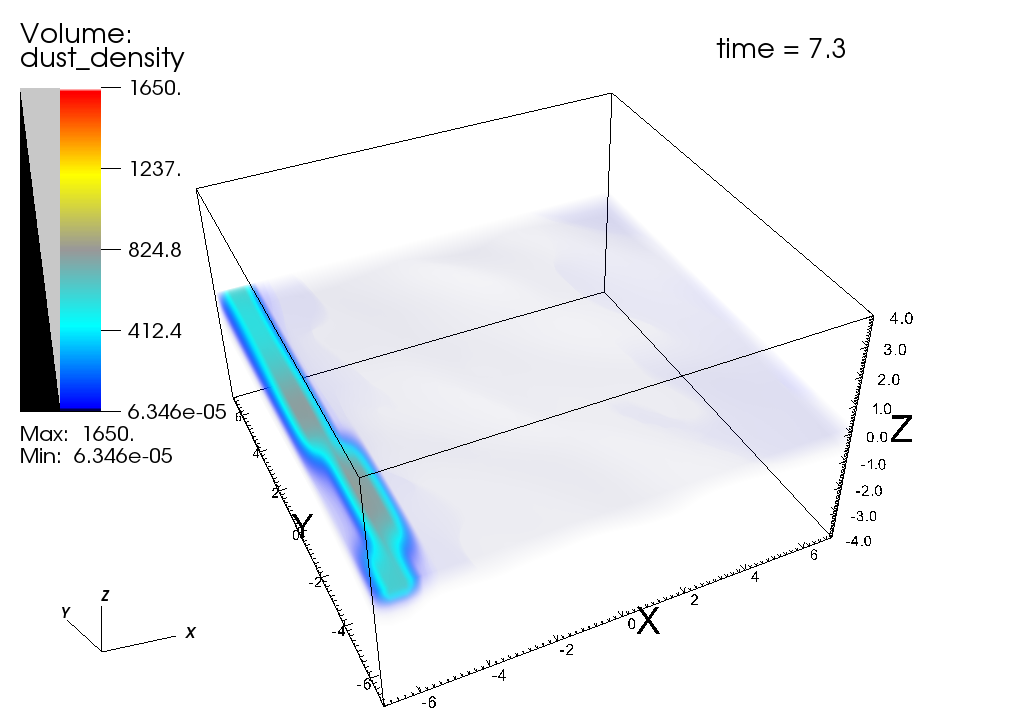}
\includegraphics[width=8cm]{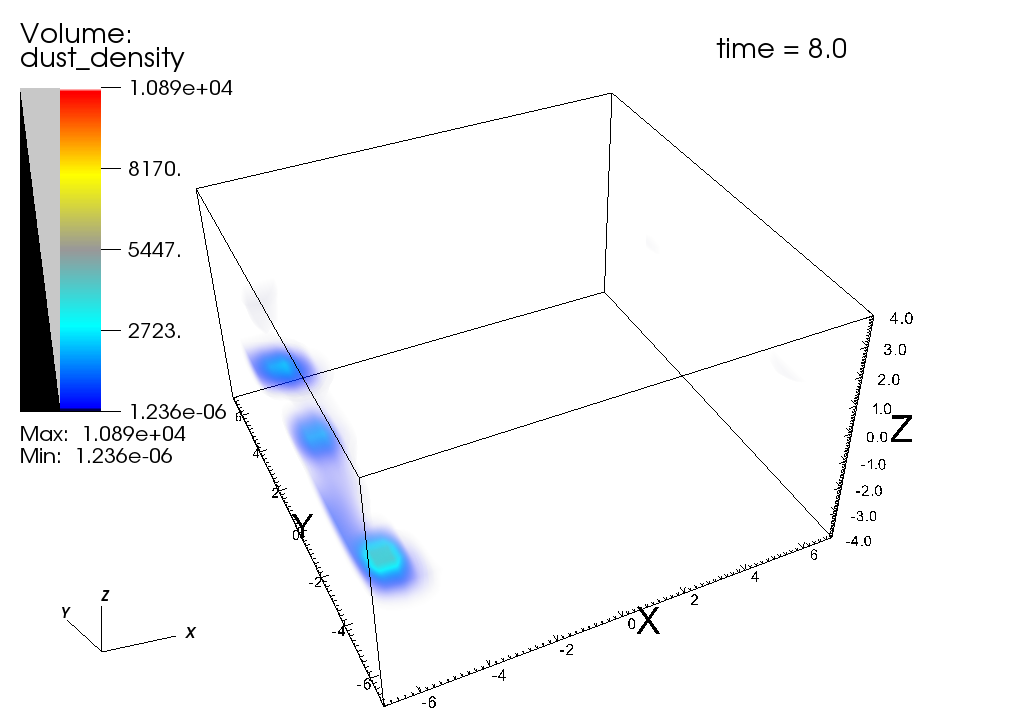}
\includegraphics[width=8cm]{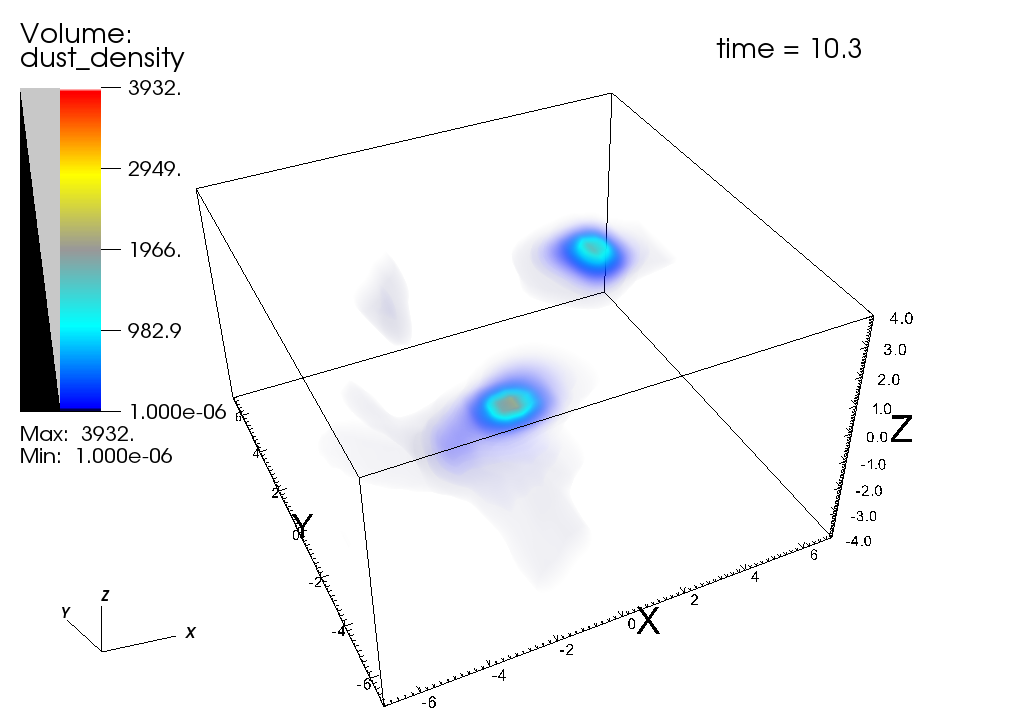}
}
\caption{\small{Time evolution of gravitational instability in our standard 3D stratified dusty disk (run S = STD32). Shown are volume renderings of dust density for the bottom half of the disk
at $t=0$, $5.0$, $6.4$, $7.3$, $8.0$, and $10.3\Omega^{-1}$ (left to right, top to bottom). 
\label{fig:vol_dust}}
}
\end{figure*}

Figure~\ref{fig:vol_dust} displays a time series of the
volume-rendered dust density in the bottom half of the box.  Over the
course of several dynamical times, density waves shear and amplify,
eventually concentrating into a single azimuthally elongated
filament. This filament then fragments radially. The fragments
gravitationally scatter and merge; by the end of the simulation, two
clumps remain.

\begin{figure}[h!]
\epsscale{1.05}
\plotone{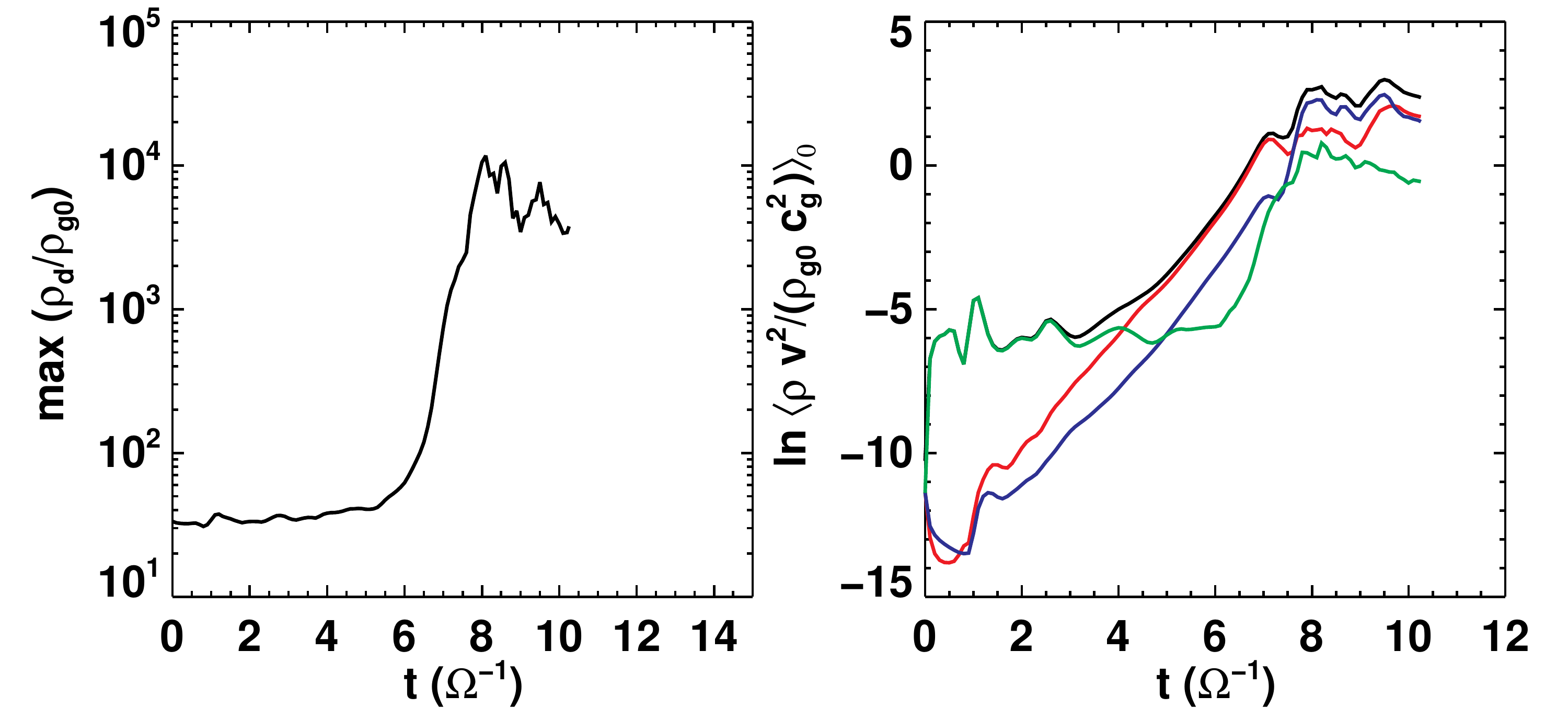}
\caption{\small{Left: Time evolution of maximum dust density for run STD32 (= S). Right: Time evolution of kinetic energies averaged horizontally and vertically over a thin slab subtending two grid cells at the midplane (red = $x$-component of kinetic energy; blue = $y$; green = $z$; black = total).
}}\label{fig:std32}
\end{figure}

A simple diagnostic that we use throughout this paper is the time
evolution of the maximum dust density, shown in the left panel of
Figure~\ref{fig:std32}. Comparison with Figure~\ref{fig:vol_dust}
reveals that $\max \rhod$ grows exponentially when the filament
fragments radially.  The maximum dust density ceases to rise once the
clumps finish coalescing. At this point each clump is gravitationally
bound,
with a maximum central density
that depends on the simulation resolution (\S\ref{sec:converge}).

The right panel of Figure~\ref{fig:std32} shows the time evolution of
various kinetic energy densities, evaluated in the three directions
and excluding the background Keplerian shear. The energy densities are
averaged horizontally and vertically over a thin slab subtending two
grid cells at the midplane (qualitatively similar results are
obtained over larger vertical averages).  The horizontal kinetic
energies grow exponentially from $t = 2$--$7 \, \Omega^{-1}$, with an
exponential growth rate of $\sim$1.5$\Omega$.
Radial motions dominate azimuthal motions until the end of the
simulation when they become comparable. Vertical motions develop
immediately after the beginning of the simulation because our
discretized initial conditions cannot be in perfect hydrostatic
balance; however the magnitude of the vertical motions is small and
stays roughly constant for $t \lesssim 6 \, \Omega^{-1}$.  For $t
\gtrsim 6 \,\Omega^{-1}$, vertical motions amplify but for the most
part remain smaller than horizontal motions.


\begin{figure}[h!]
\epsscale{1.05}
\plotone{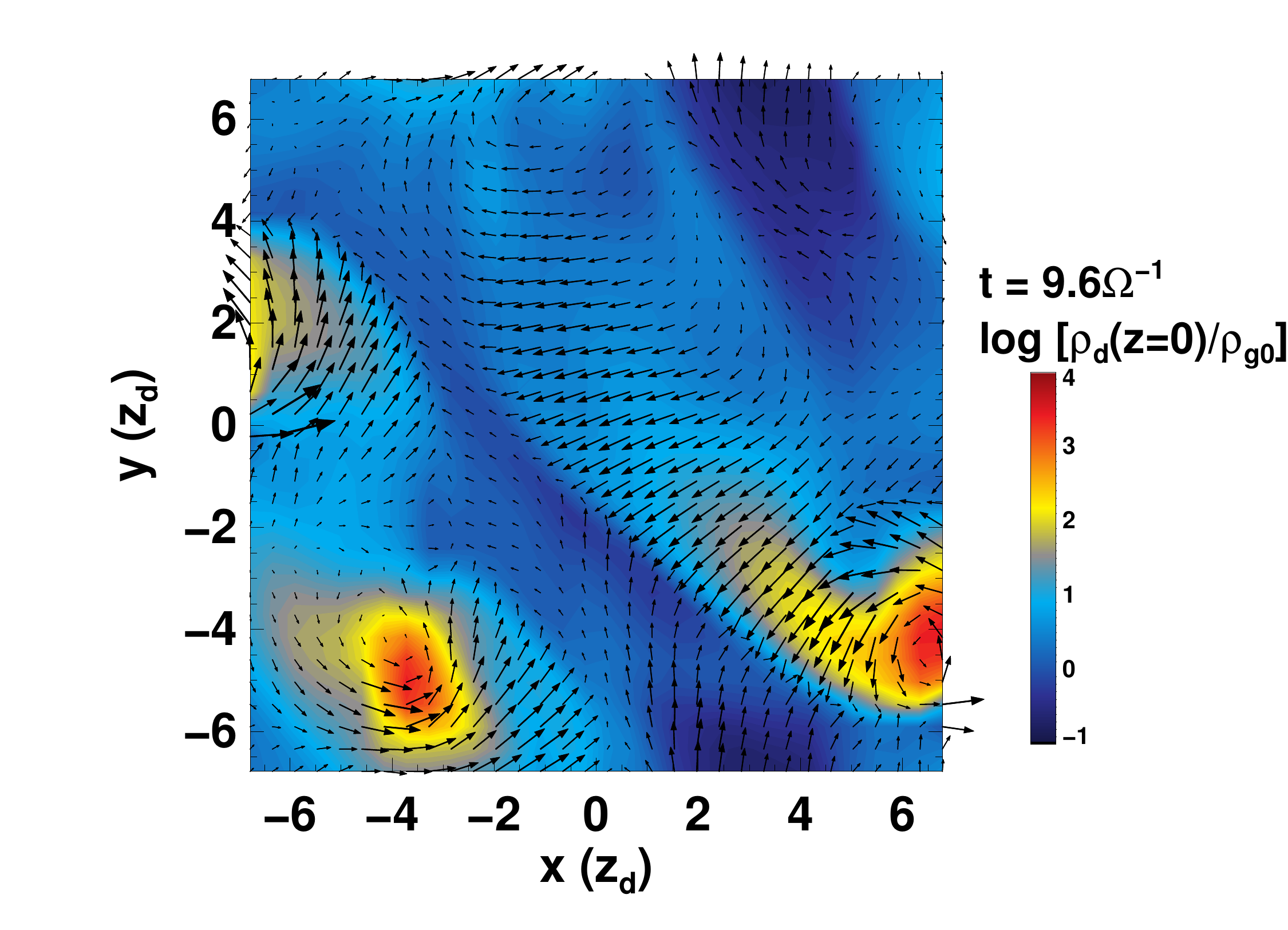} \caption{\small{Snapshot of the midplane for run S = STD32 at $t=9.6 \Omega^{-1}$. The largest in-plane velocity shown is $2.16 \, \cg$.}}\label{fig:vfield}
\end{figure}

The in-plane motions of the dusty clumps are illustrated in Figure~\ref{fig:vfield} with a snapshot
of the midplane slice of STD32 at $t=9.6\,\Omega^{-1}$. 
The dust clumps are seen spinning about their centers of mass as a consequence
of angular momentum conservation.

\subsubsection{Resolution and box size\label{sec:converge}}

Table~\ref{tab:tab4} lists the parameters of experiments designed
to test our choices for resolution, box size, and grid-cell aspect
ratio.

\begin{deluxetable}{ccccc}
\tabletypesize{\footnotesize}
\tablecolumns{5} \tablewidth{0pc}
\tablecaption{3D Simulation Parameters to Test Box Size and Resolution}
\tablehead{\colhead{Name} & \colhead{$L_x\times L_y\times L_z$($\zd^3$)} & \colhead{Resolution} & \colhead{GI\tablenotemark{(a)}} &
\colhead{~Duration ($\Omega^{-1}$)}}
\startdata
STD32(S) & $14\times14\times8$  & $32\times32\times32$     & Y & $9.8 $  \\
STD8   & $14\times14\times8$  & $8\times8\times8$        & N & $30.0$  \\
STD16  & $14\times14\times8$  & $16\times16\times16$      & Y & $11.0 $  \\
STD64  & $14\times14\times8$  & $64\times64\times64$     & Y & $11.0$    \\
U32    & $14\times14\times8$  & $56\times56\times32$     & Y & $10.0$  \\
\tableline
LZ2    & $14\times14\times2$    & $32\times32\times8$  & N & $30.0$  \\
LZ4    & $14\times14\times4$    & $32\times32\times16$ & Y & $8.5 $  \\
LZ6    & $14\times14\times6$    & $32\times32\times24$ & Y & $8.0 $  \\
LZ10   & $14\times14\times10$   & $32\times32\times40$ & Y & $9.0 $  \\
LZ14   & $14\times14\times14$   & $32\times32\times56$ & Y & $10.5$  \\
\tableline
LXY6   & $6\times6\times8$     & $16\times16\times32$  & N & $30.0$  \\
LXY10  & $10\times10\times8$   & $24\times24\times32$  & Y & $8.6 $  \\
LXY20  & $20\times20\times8$   & $48\times48\times32$  & Y & $8.7 $  \\

\enddata \label{tab:tab4}
\tablenotetext{(a)}{~GI = Gravitational Instability. Y means $\max \rhod$ increases by orders of magnitude over a few dynamical times, and N means it does not.}
\end{deluxetable}

\begin{figure}[h!]
\plotone{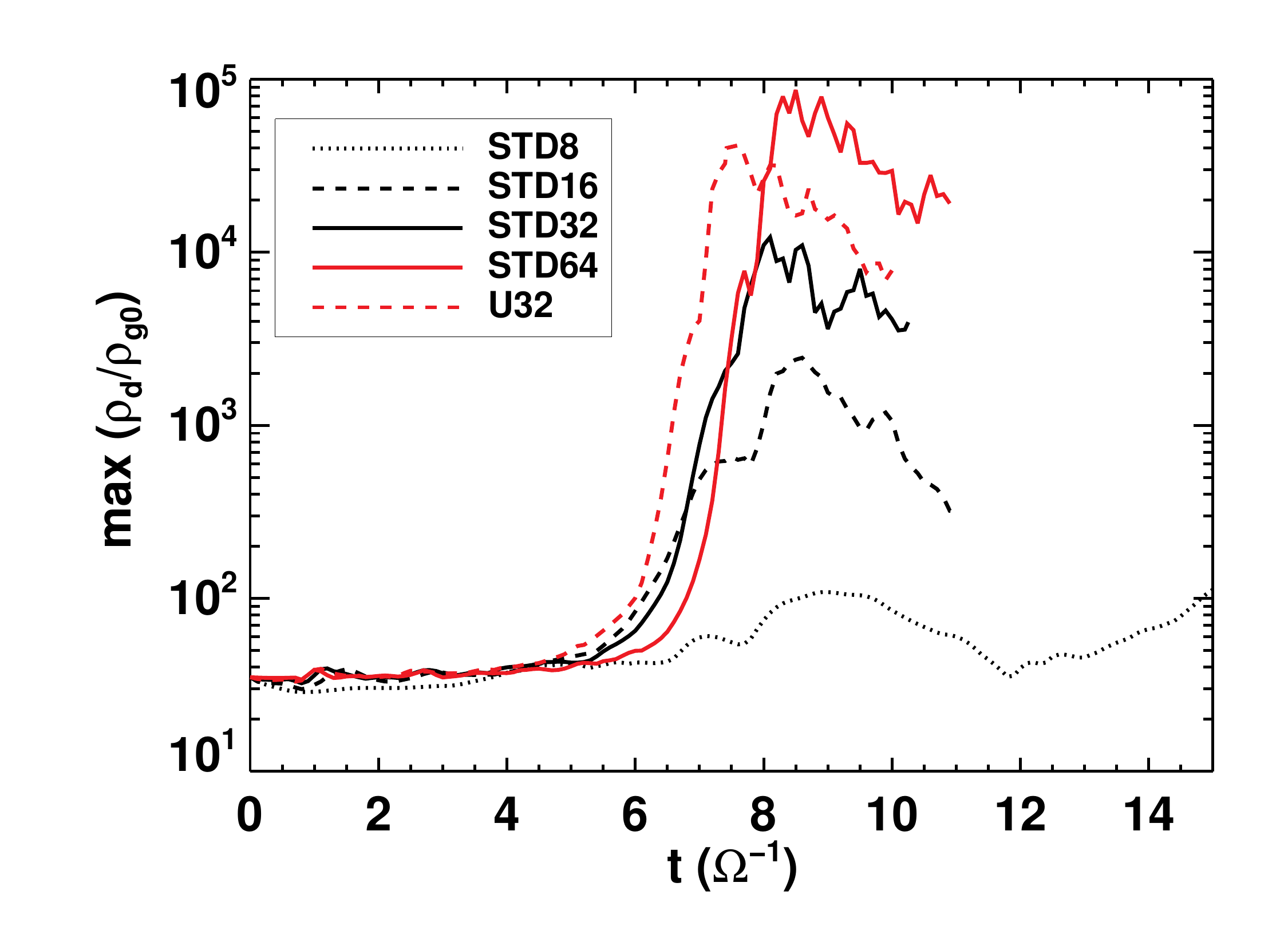} \caption{\small{Time evolution of the maximum dust density for the resolution
study (see Table \ref{tab:tab4}).}}\label{fig:dmax_resol}
\end{figure}

Figure~\ref{fig:dmax_resol} shows how varying the resolution changes
the evolution of our standard, gravitationally unstable run
(\textrm{STD32} --- also labeled \textrm{S} in Table~\ref{tab:tab3}).
We use again the simple metric of $\max \rhod$ vs.~$t$.  Broadly
speaking, the runs \textrm{STD16}, \textrm{STD32}, \textrm{STD64} are
all ``acceptable'' insofar as they all yield increases in $\max \rhod$
by orders of magnitude within several dynamical times ($t \lesssim 8
\,\Omega^{-1}$). 
By contrast, the lowest resolution run, \textrm{STD8}, is unacceptable.
%
Thus, the minimum acceptable
resolution appears to be $\sim$2 cells per scale length $\zd$ in the
vertical direction (cf.~\citealt{nelson_a2006} who found that a minimum of four smoothing lengths per scale height is required for SPH simulations),
and $\sim$8 cells per critical wavelength
$\lambda_c$ in the horizontal directions. Our standard choices for
resolution --- as well as the resolutions characterizing all our
``science'' runs, listed in Table~\ref{tab:tab3} and discussed in
\S\ref{sec:parameter} --- satisfy these minimum requirements by a
safety factor of 2.

Examining Figure \ref{fig:dmax_resol} more critically, we see that the
maximum value attained by $\max \rhod$ has not converged with
resolution. Increasing the resolution enables us to resolve ever
higher densities in the collapsing clumps. Another point of concern is
the non-uniform aspect ratios of individual grid cells, which ranges
from $x$:$y$:$z$ $\approx$ $2$:$2$:$1$ to $4$:$4$:$1$ over our set of science
simulations (Table \ref{tab:tab3}).  The run \textrm{U32} is
characterized by perfectly cubical grid cells ($1$:$1$:$1$); the
evolution is similar to \textrm{STD32}, but is characterized by an
earlier onset of gravitational instability, and stronger density
fluctuations.  This comparison suggests that our science runs
with non-cubical grid cells are biased slightly against gravitational
instability.

\begin{figure}[h!]
\plotone{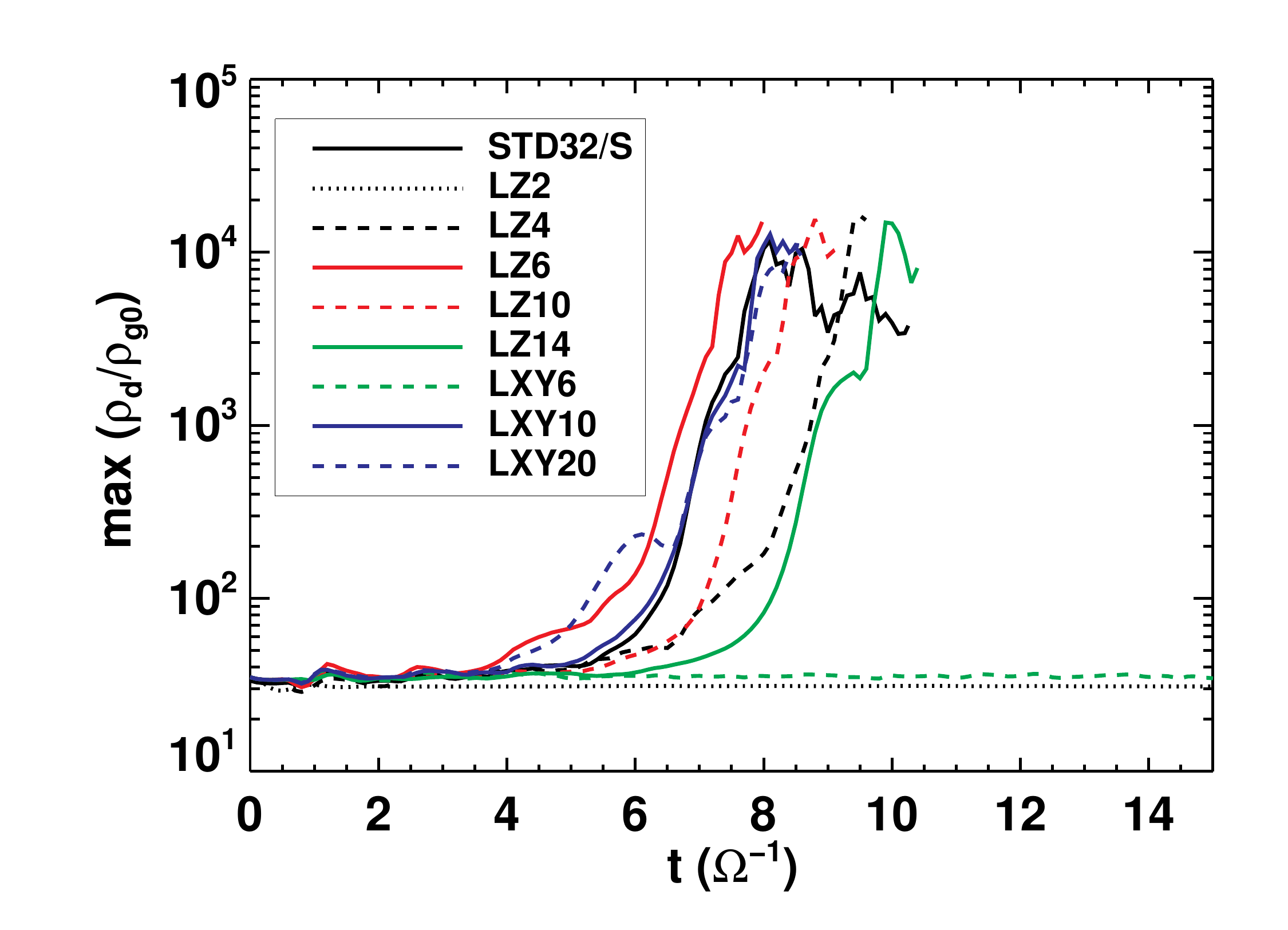} \caption{\small{Time evolution of the maximum dust density for our box
size tests (see Table \ref{tab:tab4}).}}\label{fig:dmax_size}
\end{figure}

We next investigate how box size affects our results. For all box size
experiments, the spatial resolution is kept at its standard value (32
grid cells per 14$\zd$ in either horizontal direction, and 4 grid
cells per $\zd$ in the vertical direction).  Runs \textrm{LZ2} through
\textrm{LZ14} vary box height $L_z$ while keeping $L_x$ and $L_y$
fixed at their standard (STD32 = S) values.  As
Figure~\ref{fig:dmax_size} reveals, box heights of $4$--$14\zd$ yield
comparable results, while a box height of $2\zd$ is unacceptable.  For
the most part, increasing the box height seems to delay the onset of
gravitational instability, with \textrm{LZ4} being the exception to
this rule.

Our 2D simulations indicated that $L_x$ and $L_y$ must be large enough
to encompass at least one critical wavelength $\lambda_c$. Our 3D
simulations bear out this same requirement. 
Figure~\ref{fig:dmax_size} shows that run \textrm{LXY6}, for which
the box size is just under one critical wavelength,
does not exhibit gravitational instability, unlike its bigger
box counterparts.

To summarize our findings in this subsection: (1) The simulation box
should be at least $4\zd$ tall ($2\zd$ above and below the
midplane). (2) Each horizontal dimension must be longer than one
critical wavelength $\lambda_c$ as given by equation
(\ref{eq:critical_wavelength}). (3) Simulations require a vertical
resolution of $\gtrsim 2$ grid cells per scale length $\zd$, and a
horizontal resolution of $\gtrsim 8$ grid cells per critical
wavelength. (4) Individual grid cells that have increasingly
non-uniform aspect ratios (squatter vertically than horizontally)
tend to suppress gravitational instability, but the bias is minor
and aspect ratios up to 4:4:1 appear acceptable.
All of our science simulations (Table~\ref{tab:tab3};
\S\ref{sec:parameter}) satisfy these requirements, in some cases
by factors of 2.

\subsubsection{Criteria for gravitational collapse \label{sec:parameter}}

\begin{figure}[h!]
  \plotone{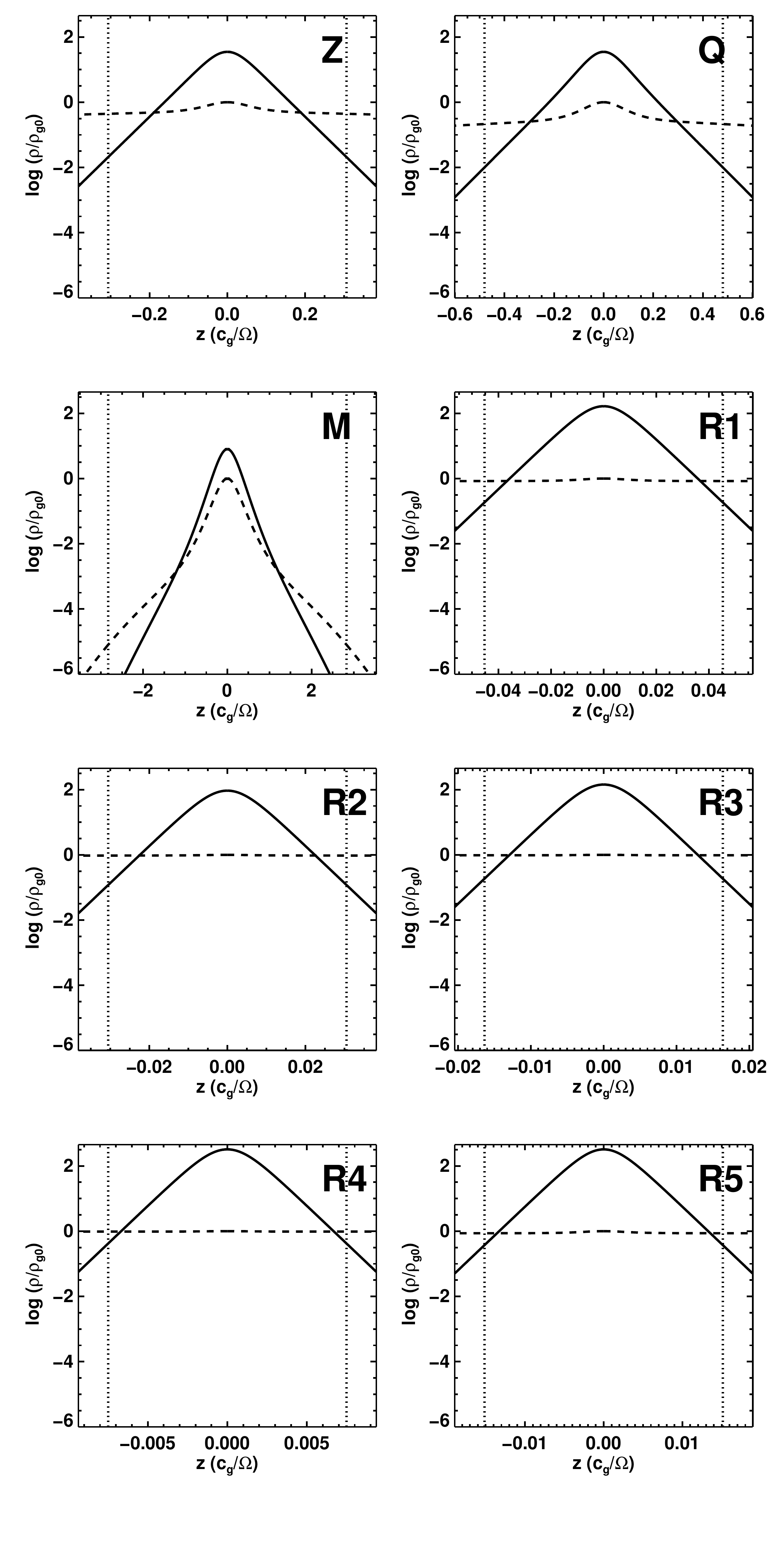} \caption{\small{Initial conditions for our
      science runs which explore parameter space. Solid lines denote
      dust, and dashed lines denote gas. The vertical lines delimit
      the vertical boundaries of our simulation box.
    }}\label{fig:init8}
\end{figure}

\begin{figure}[h!]
  \includegraphics[width=\linewidth]{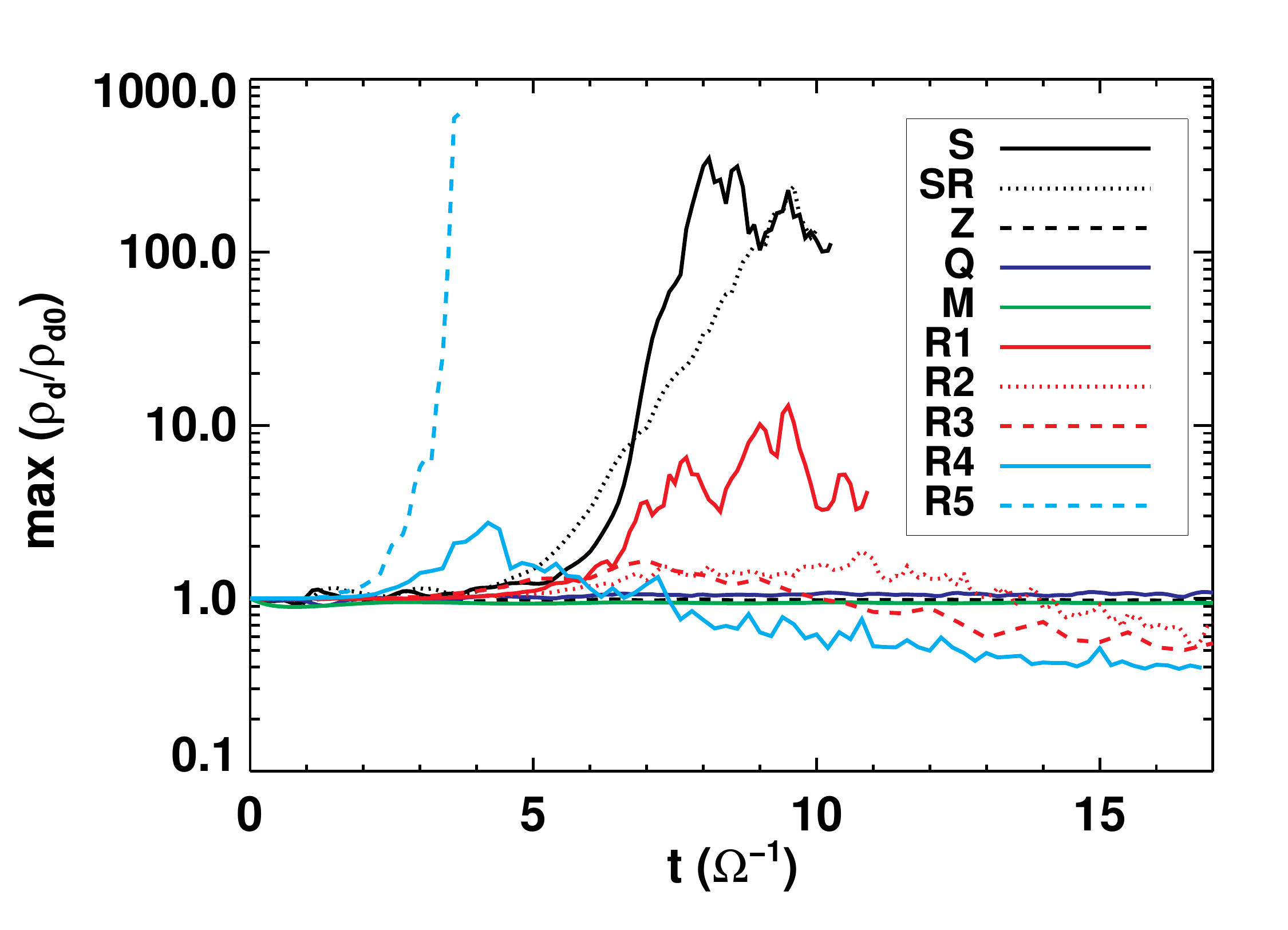} \caption{\small{Time
      evolution of the maximum dust density in our science
      simulations. Only for runs S, SR, and R5 does $\rho_0 > \rho_{\rm
        II}^\ast$, and indeed only those runs exhibit dramatic growth
      of the dust density due to gravitational instability.
    }}\label{fig:dmax_3d}
\end{figure}

Table \ref{tab:tab3} lists the simulations
designed to test which of the various proposed criteria
for gravitational instability is the best predictor of collapse.
Figures \ref{fig:init0} and \ref{fig:init8} describe
the initial dust and gas profiles, while Figure \ref{fig:dmax_3d}
displays the results using our simple diagnostic
of $\max \rhod$ vs.~time.

First consider runs S and R1--R5, and ask whether these runs
favor $\rhoone$ or $\rhotwo$ for the density
required for gravitational collapse.  Because dust is a major
component of our disks, we do not expect $\rhoone$ --- which
is strictly valid only for pure gas disks --- to be a good
predictor. Indeed in all six of these runs, the midplane
density $\rho_0$ exceeds $\rhoone$, by factors of
7.5--30, yet only runs S and R5, and to a much lesser extent R1, exhibit
collapse.  All six runs indicate instead that $\rhotwo$
--- equivalently, $\Qd$ --- is the better predictor, with the critical
value
\begin{equation} \label{eq:qdstar}
0.5 < Q_{\rm d}^\ast < 0.9 \,.
\end{equation}
There is some concern that the comparison between
runs R2--R5 and run S may not be fair because
runs R2--R5 have a factor of $\sim$2
poorer spatial resolution in $x$ and $y$ compared to run S.
This concern is allayed by run SR, which has the same
physical parameters as S but is run with the box size
and resolution of R3, and which turns out to
behave qualitatively similarly to S (see Figure \ref{fig:dmax_3d}).

Our conclusion that $\rhotwo$ is relevant
and that $\Qd^\ast$ obeys (\ref{eq:qdstar})
is supported further by runs Z, Q, and M,
each of which varies one of the three input parameters
$\metal$, $\Qd$, and $\mu_0$.

\begin{figure}[h!]
\epsscale{1.10}
\plotone{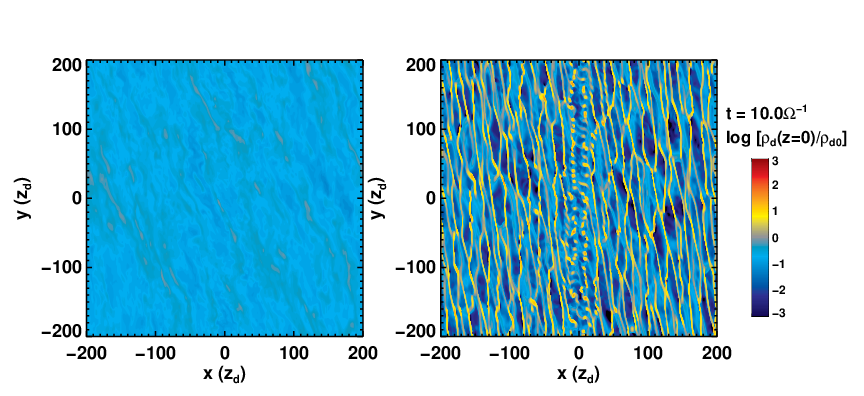} 
\caption{\small{Snapshots of the dust density at the midplane for run R3 (left panel) and run SR (right panel). In R3, the initial midplane density $\rho_0 > \rho_{\rm Sekiya}^\ast$, and there is some clumping, but it
is much lower in amplitude compared to run SR, for which $\rho_0 > \rhotwo$.}}\label{fig:compare}
\end{figure}

Although runs R2--R4 do not exhibit the dramatic growth in $\rhod$
shown by runs S, SR, and R5 --- a result that we interpret to mean that
$\rhotwo$ gives the correct criterion for gravitational
collapse --- runs R2--R4 do show some clumping.  Figure
\ref{fig:compare} compares snapshots of runs R3 and SR (performed with
the same box size and resolution), taken at the same time $t =
10\Omega^{-1}$.  Filaments do form in R3, although they are much
weaker in density contrast compared to the filaments in SR. The mild
growth shown in runs R2 and R3 might simply reflect the fact that
their values for $\Qd = 1.86$ are still too close to $\Qd^\ast$ to
suppress instability entirely.  
An alternative (and not mutually
exclusive) possibility is that because $\rho_0 > \rho_{\rm
  Sekiya}^\ast = 0.60 \rho^\dagger$ for runs R2--R4, the disk might
be exhibiting the unstable (and formally incompressible) mode found by
\citet{sekiya83}.  
Whatever the interpretation, the modest growth
factors exhibited by R2--R4 seem unlikely to lead to planetesimal formation.
In particular, the density concentrations in runs R2--R4 eventually
disperse, unlike the density concentrations in runs S, SR, and R5 for
which $\rho_0 > \rhotwo$. What evidence we have suggests
that Sekiya's mode is not important for planetesimal formation,
but higher resolution simulations that better
separate $\rho_{\rm Sekiya}^\ast$ from $\rhotwo$ are needed for a more
definitive assessment.

Finally, what about $\rho^\ast_{\rm Roche}$ vs.~$\rhotwo$? Here runs
R4 and R5 are the most telling. Both runs are characterized by the
largest midplane densities $\rho_0 > \rho_{\rm Roche}^\ast$, but only
R5, for which $\rho_0 > \rhotwo$, undergoes gravitational collapse
(see Figure \ref{fig:dmax_3d}).

Table \ref{tab:tab5} summarizes
how the various candidate critical densities relate
to one another and to the midplane density for our science simulations.
From Table \ref{tab:tab5}, $\rhotwo$ emerges as the best predictor of
collapse.

\begin{deluxetable}{cr@{}lc}
\tabletypesize{\footnotesize}
\tablecolumns{4} \tablewidth{0pc}
\tablecaption{Comparison of Critical Densities and Actual Midplane Density for Science Simulations}
\tablehead{\colhead{Name} & \multicolumn{2}{c}{Critical density relations} & \colhead{GI\tablenotemark{(a)}}} 
\startdata
\textrm{S}  & $\rhoone<\rhotwo<\rho_{\rm Sekiya}^\ast<$ &~$\rho_0<\rho_{\rm Roche}^\ast$ & Y \\
\textrm{R1} & $\rhoone<\rho_{\rm Sekiya}^\ast<\rhotwo<$ &~$\rho_0<\rho_{\rm Roche}^\ast$ & Y/N\tablenotemark{(b)} \\
\textrm{R2} & $\rhoone<\rho_{\rm Sekiya}^\ast<        $ &~$\rho_0<\rho_{\rm Roche}^\ast<\rhotwo$ & N \\
\textrm{R3} & $\rhoone<\rho_{\rm Sekiya}^\ast<        $ &~$\rho_0<\rho_{\rm Roche}^\ast<\rhotwo$ & N \\
\textrm{R4} & $\rhoone<\rho_{\rm Sekiya}^\ast<\rho_{\rm Roche}^\ast<        $ &~$\rho_0<\rhotwo$ & N \\
\textrm{R5} & $\rhoone<\rho_{\rm Sekiya}^\ast<\rhotwo<\rho_{\rm Roche}^\ast<$ &~$\rho_0$ & Y \\
\enddata \label{tab:tab5}
\tablenotetext{(a)}{~GI = Gravitational Instability. Y means $\max \rhod$ increases by orders of magnitude over a few dynamical times, and N means it does not.}
\tablenotetext{(b)}{~See Figure \ref{fig:dmax_3d}.}
\end{deluxetable}

\section{SUMMARY AND DISCUSSION \label{sec:discuss}}

\begin{figure}[h!]
  \plotone{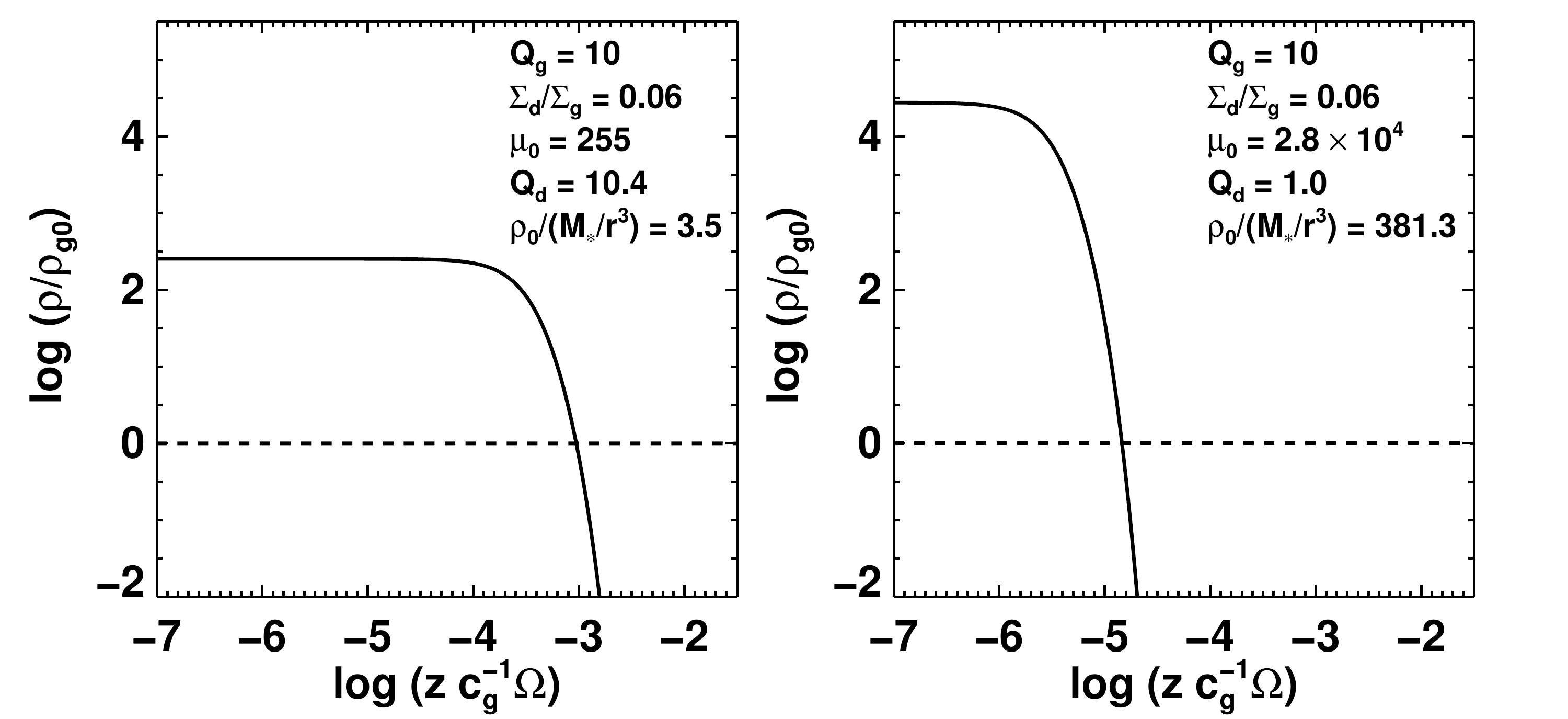} \caption{\small{A tale of two particle sublayers, one
      of which is thinner and denser than the
      other. Dust density is plotted as a solid line, and gas density
      as a dashed line. The disks have identical masses and bulk
      metallicities, enhanced over those of the
      minimum-mass solar nebula by factors of 3--4.  Left: Midplane
      density $\rho_0 = \rho^\ast_{\rm Roche} = 3.5 \rho^\dagger$ and
      $\Qd = 10.4$.  Right: Midplane density $\rho_0 \approx \rhotwo \approx
      10^2 \rho^\ast_{\rm Roche}$ and $\Qd = 1$.  According to the
      results of our simulations, only the model in the right panel,
      having the thinner and denser sublayer, should be on the verge
      of gravitational collapse --- in the limit that particles are
      aerodynamically perfectly coupled to gas. We argue in
      \S\ref{sec:wayout} that when the perfect coupling approximation
      breaks down, it may be possible for the disk on the left to
      undergo gravitational instability.}}\label{fig:init_physical}
\end{figure}

Dust grains settle toward the midplanes of protoplanetary disks, forming
a sublayer of solid particles sandwiched from above and below by gas.
Whether this sublayer can become thin enough and dense enough to
undergo gravitational instability and fragment into planetesimals
is an outstanding question. We have found in this work that the density
threshold for gravitational collapse can be extraordinarily high --- much
higher even than the Roche density $\rho^\ast_{\rm Roche} = 3.5 M_\ast / r^3$,
where $M_\ast$ is the mass of the central star and $r$ is the orbital
radius. To trigger collapse in the limit that dust particles
are small enough to be tightly coupled to gas, the density $\rho_0$
in the sublayer must be such that the Toomre stability
parameter
\beq
\Qd \approx \left(\frac{\rhotwo}{\rho_0}\right)^{1/2} \lesssim 1
\enq
where
\beq
\rhotwo \approx \frac{M_{\ast}}{2\pi r^3} \frac{\Qg}{\Qd^{\ast 2}} \left( \frac{\Sigmag}{\Sigmad} \right)^2 \,.
\enq
(For more precise relations, see equations \ref{eqn:Q_d}, \ref{eqn:rho00}, and \ref{eq:qdstar}.) Here $\Qg$ is the Toomre parameter for the ambient (and much thicker) gas
disk, $\metal$ is the ratio of surface densities of dust and gas (i.e., the height-integrated metallicity),
and $0.5 < \Qd^{\ast} < 0.9$ as measured from our simulations.
For an astrophysically plausible disk having $3\times$ the mass
of the minimum-mass solar nebula ($\Qg \approx 10$) and a bulk
metallicity enriched over solar by a factor of 4 ($\metal \approx 0.06$),
the critical density
\beq
\rhotwo \approx 1.3\times 10^2 \Qd^{\ast -2} \rho_{\rm Roche}^\ast \,.
\enq

Figure \ref{fig:init_physical} portrays two sublayers --- one for
which $\rho_0 = \rho^\ast_{\rm Roche}$ and another, much thinner
sublayer for which $\rho_0 \approx \rhotwo \approx 10^2 \rho^\ast_{\rm
  Roche}$ ($\Qd = 1$). The results of our simulations, performed in
the limit of perfect aerodynamic coupling between particles and gas,
indicate that only the latter, much denser disk is on the verge of fragmenting.

Qualitatively, such extraordinary densities are required for gravitational
instability because gas pressure renders the sublayer extremely stiff.
Sound-crossing times for thin layers are easily shorter than free-fall times.
We can examine the competition between stabilizing pressure,
stabilizing rotation, and de-stabilizing self-gravity in both the horizontal
(in-plane) and vertical directions. Horizontal stability
is controlled by $\Qd$: when $\Qd > \Qd^\ast \sim 1$, all horizontal
lengthscales $\lambda \lesssim 2\cd^2/G\Sigmad$ are stabilized
by pressure, and all scales $\lambda \gtrsim 2\cd^2/G\Sigmad$ are stabilized
by rotation, where $\cd$ is the effective sound speed in the dust-gas mixture.
At the same time, vertical stability is assured whenever the sound-crossing
time across the vertical thickness of the sublayer $2H_{\rm d}$
is shorter than the free-fall time:
\beq
\frac{2H_{\rm d}}{\cd} < \frac{1}{\sqrt{G\rhod}}
\enq
which, after substituting $\Hd \approx \Sigmad / 2 \rhod$ and $\cd
\approx \cg \sqrt{\rhog/\rhod}$, translates to
\beq
\left( \frac{\Sigmad}{\Sigmag} \right)^2 \frac{1}{\Qg} < \frac{\pi}{2}
\enq
which is easily satisfied for reasonable disk parameters.

The severe obstacle that gas pressure presents to gravitational collapse
of aerodynamically well-coupled particles is discussed
by Cuzzi, Hogan, \& Shariff (\citeyear{cuzzietal08}, see their section 3.1).
Our 3D disk simulations support their 1D considerations.

\subsection{Directions for Future Research}\label{sec:wayout}

Taken at face value, the higher density threshold $\rhotwo$ established by our
work argues against using aerodynamically well-coupled
particles to form planetesimals.  The Kelvin-Helmholtz instability
(KHI) may prevent dust from settling into the extraordinarily thin
sublayers needed to cross the density threshold. One potential loophole
is provided by \citet{sekiya98} and \citet{youdinshu02}, who found in 1D
that self-gravitating, non-rotating sublayers having constant
Richardson number $Ri$ could develop cusps of infinite density at the
midplane. The presumption of these studies is that dust settles into a
state that is marginally KH-stable and that this state is
characterized by a constant $Ri$.  Some evidence for a spatially
constant $Ri$ was found in the settling experiments of \citet{leeetal10b},
but only near the top and bottom faces of the dust
sublayer and not at the midplane.  These numerical experiments
suffered, however, from lack of spatial resolution toward the
midplane, and moreover neglected self-gravity.  Future simulations
of cuspy dust profiles including self-gravity would be welcome.

We have worked in the limit that the stopping times $t_{\rm stop}$
of particles in gas are small compared to all other timescales.
But in reality, finite particle sizes imply finite $t_{\rm stop}$
(see Figure \ref{fig:tstop}). When the assumption of infinitesimal
stopping time breaks down, new effects may appear that might lower
the threshold for gravitational instability.

One such effect is as follows. Consider again the competition
between stabilizing pressure and de-stabilizing self-gravity (in either the vertical or horizontal directions). A major reason why the sublayer so strongly resists collapse
is that sound waves travel quickly across it. We have taken the sound
speed for our dust-gas suspension to be $\cd = \cg / \sqrt{1 + \rhod/\rhog} \approx \cg \sqrt{\rhog/\rhod}$ (equations \ref{eqn:eos} and \ref{eqn:c_d}).
But this presumes that particles are perfectly coupled to gas. If the sound-crossing time across some scale $\lambda$ were to become shorter
than the particle stopping time, i.e., if
\beq \label{eqn:wayout}
\frac{\lambda}{\cd} \approx \frac{\lambda}{\cg} \sqrt{ \frac{\rhod}{\rhog} } < t_{\rm stop}
\enq
then our use of $\cd \approx \cg \sqrt{\rhog/\rhod}$ would be invalid.
Particles on scales $\lambda$ would lose support from gas pressure
and become susceptible to gravitational instability.

To get a sense of where in parameter space this instability may lie,
we normalize $\lambda$ to the full vertical thickness of the sublayer:
\beq
\lambda \equiv 2 \Hd \hat{\lambda}= \frac{\hat{\lambda} \Sigmad}{\rhod} \,.
\enq
where $\hat{\lambda}$ can take any value (larger than or smaller than unity). Then equation (\ref{eqn:wayout}) for the loss of pressure
support translates to a midplane density (dominated by dust) of
\beq \label{eqn:wayout_1}
\rho_0 \approx \rhod \gtrsim \frac{2}{\pi} \frac{M_\ast}{r^3} \left( \frac{\Sigmad}{\Sigmag} \right)^2 \frac{\hat{\lambda}^2}{\Qg} \frac{1}{ \left( \Omega t_{\rm stop} \right)^2} 
\enq
where $\Omega$ is the Kepler orbital frequency.
For self-gravity to resist tidal disruption, $\rhod = \rho^\ast_{\rm Roche} = 3.5 M_\ast / r^3$. Substituting this requirement into (\ref{eqn:wayout_1}), we find
that
\begin{align}
\Omega t_{\rm stop} & \gtrsim \left( \frac{2}{3.5\pi} \right)^{1/2} \left( \frac{\Sigmad}{\Sigmag} \right)
\frac{\hat{\lambda}}{\Qg^{1/2}} \nonumber \\
 & \gtrsim 8 \times 10^{-3} \left( \frac{\metal}{0.06} \right) \left( \frac{\hat{\lambda}}{1}
 \right) \left( \frac{10}{\Qg} \right)^{1/2}
\label{eqn:wayout_2}
\end{align}
for particles on scales $\hat{\lambda}$ to decouple from sound waves.
For $\hat{\lambda} = 1$, requirement (\ref{eqn:wayout_2}) could be fulfilled by particles having sizes of a few millimeters to a few centimeters at distances
of 1--10 AU (Figure \ref{fig:tstop} --- but note that the curves in the figure need to be adjusted by factors of a few for mass-enriched nebulae). For $\hat{\lambda} < 1$, even smaller particles could lose pressure support and collapse gravitationally.

Future simulations that include finite particle stopping times
could try to find such an instability. A complication would be that
accounting for finite $t_{\rm stop}$ would introduce the streaming
instability, which could prevent the dust density from attaining
the Roche value --- see, e.g., runs R21-3D and R41-3D in Figure 5
of \citet{baistone10}, for which $\Omega t_{\rm stop} \leq 0.1$
and $\rhod < \rho^{\ast}_{\rm Roche}$. To find the instability
that we are envisioning, one would have to restrict $\Omega t_{\rm stop}$
to small enough values to suppress the streaming instability --- thereby
permitting the setting of grains into sublayers for which
$\rhod = \rho^{\ast}_{\rm Roche}$ --- while at the same time
keeping $\Omega t_{\rm stop}$ large enough to satisfy (\ref{eqn:wayout_2})
and nullify pressure support.

\acknowledgments

We are grateful to Eve Ostriker and Andrew Youdin for
discussions. Section \ref{sec:wayout} was inspired by discussions with
Eve that clarified the limitations of our study and pointed the way to
a possible new route to gravitational instability.  We thank Xuening
Bai, Chang-Goo Kim, Eve Ostriker, Ian Parrish, and Jim Stone for help
in augmenting \texttt{Athena}. The simulations were performed with the
Berkeley cluster Henyey, which was made possible by a National Science
Foundation Major Research Instrumentation (NSF MRI) grant.  Financial
support for the authors was provided by the Berkeley Center for
Integrative Planetary Science, the Berkeley Theoretical Astrophysics
Center, and grants from NSF (AST-0909210) and NASA Origins.


\bibliography{shear}



\end{document}